\documentclass[twocolumn,showpacs,preprintnumbers,amsmath,amssymb]{revtex4}

\usepackage{graphicx}
\usepackage{dcolumn}
\usepackage{bm}

\newcommand{\epsfigbox}[5]{%
\begin{figure} \vspace{#3}%
\includegraphics[width=7.0cm]{#2}%
\caption{ \label{fig:#1} #5}
\vspace{#4}
\end{figure}}

\newcommand{\tsub}[1]{_{\mbox{\scriptsize#1}}}

\newcommand{\quarterthin}{\kern 0.0417em}

\newcommand{\bra}[1]{\langle#1|}
\newcommand{\ket}[1]{|#1\rangle}

\newcommand{\atanh}{{\rm atanh}}

\newcommand{\DeltaZero}{\Delta_0}

\newcommand{\Deltad}{\Delta_d}
\newcommand{\Deltaq}{\Delta_q}
\newcommand{\Deltapi}{\Delta_\pi}

\newcommand{\Deltadsq}{\Delta_d^2}
\newcommand{\Deltaqsq}{\Delta_q^2}
\newcommand{\Deltapisq}{\Delta_\pi^2}
\newcommand{\Deltapm}{\Delta_\pm}
\newcommand{\DeltaMinus}{\Delta_-}
\newcommand{\DeltaPlus}{\Delta_+}
\newcommand{\lambdaPrime}{\lambda^\prime}

\begin{document}


\title
{Temperature-dependent gap equations and their solutions \\
in the SU(4) model of high-temperature superconductivity}

\author{Yang Sun$^{(1)}$}
\email{ysun@nd.edu}
\author{Mike Guidry$^{(2)}$}
\email{guidry@utk.edu}
\author{Cheng-Li Wu$^{(3)}$}
\email{clwu@phys.cts.nthu.edu.tw}

\affiliation{$^{(1)}$Department of Physics, University of Notre
Dame, Notre Dame, Indiana 46556, USA \\
$^{(2)}$Department of Physics and Astronomy, University of
Tennessee, Knoxville, Tennessee 37996, USA \\
$^{(3)}$Physics Division, National Center for Theoretical Science,
Hsinchu, Taiwan 300, ROC}

\date{\today}

\begin{abstract}
Temperature-dependent gap equations in the SU(4) model of
high-$T\tsub c$ superconductivity are derived and analytical
solutions are obtained. Based on these solutions, a generic gap
diagram describing the features of energy gaps as functions of
doping $P$ is presented and a phase diagram illustrating the phase
structure as a function of temperature $T$ and doping $P$ is
sketched. A special doping point $P_q$ occurs naturally in the
solutions that separates two phases at temperature $T=0$: a pure
superconducting phase on one side ($P>P_q$) and a phase with
superconductivity strongly suppressed by antiferromagnetism on the
other ($P<P_q$). We interpret $P_q$ as a quantum phase transition
point. Moreover, the pairing gap is found to have two solutions
for $P<P_q$: a small gap that is associated with competition
between superconductivity and antiferromagnetism and is
responsible for the ground state superconductivity, and a large
gap without antiferromagnetic suppression that corresponds to a
collective excited state. A pseudogap appears in the solutions
that terminates at $P_q$ and originates from the competition
between $d$-wave superconductivity and antiferromagnetism.
Nevertheless, this conclusion does not contradict the preformed
pair picture conceptually if the preformed pairs are generally
defined as any pairs formed before pairing condensation.
\end{abstract}


\maketitle

\section{Introduction}

The properties of of high-temperature (high-$T\tsub c$)
superconductivity in copper oxide materials pose a strong
challenge to the theoretical understanding of superconductivity in
strongly correlated many-body systems. The main question
concerning the high-$T\tsub c$ superconductivity phase diagram is
the transition between the antiferromagnetic (AF) and
superconducting (SC) phases, which is dominated by anomalous
properties commonly attributed to a pseudogap in the spectrum
\cite{Ti99}. Partly because of the complicated material properties
and the nature of the experiments performed, different experiments
seem to emphasize different aspects of the system, leading to
different and sometimes contradictory possibilities for
explanation of observables \cite{Shen03}. Many models have been
proposed in the last two decades but there is no consensus as to
the origin of the pseudogap properties.  Still, one cannot answer
the question: what is the real phase diagram in cuprates?

Describing collective motion in a strongly correlated many-body
system in terms of single particle degrees of freedom is not
always feasible. In the cuprate systems exhibiting high-$T\tsub c$
superconducting properties, there is a substantial point of view
(see, for example, Ref.\ \cite{An00}) that the many-body
correlations are so strong that the dynamics can no longer be
described meaningfully in terms of individual fundamental
particles. However, collective motions in such quantum many-body
systems are often governed by only {\it a few} collective degrees
of freedom. Once these degrees of freedom are identified and
properly incorporated into a model, calculations may become
feasible and, more importantly, the physics may become
transparent. One systematic approach to this program of
identifying the relevant collective degrees of freedom in a
many-body system is the method of dynamical symmetries.

Examples of using the dynamical symmetry approach to study
many-body problems may be found in nuclear physics
\cite{FDSM,IBM}, molecular physics \cite{Mole1,Mole2}, and
particle physics \cite{Part}. Recently, we have developed an SU(4)
dynamical symmetry model \cite{Gu99,lawu99,wu03} aimed at
understanding the mechanism that leads to high-$T\tsub c$
superconductivity in cuprates. Theoretical models employing
algebras and groups have been used in condensed matter physics
(for recent examples, see Refs. \cite{Examples,Zh97}). However, to
our knowledge, the powerful dynamical symmetry methods that we
employ here have not been applied systematically in this field.

In this paper we derive and solve the temperature-dependent gap
equations in the SU(4) model. We demonstrate that this model can
provide important insight into the puzzling issues associated with
cuprate superconductors. In the next section we outline the SU(4)
model (a more detailed description can be found in Refs.
\cite{Gu99,lawu99,wu03}). Temperature-dependent gap equations of
the SU(4) model are derived in section III. Analytical solutions
for the gap equations at $T=0$ are obtained in section IV, and a
generic gap diagram (energy gaps versus doping at $T=0$) predicted
by the SU(4) model is presented. The solutions of the gap
equations at finite temperature are given in section V and these
are used to construct a generic phase diagram in section VI.
Finally, a summary is given in section VII. Technical details
concerning construction, derivation, and solution of the gap
equations are given in three appendixes.

\section{The SU(4) model}

It is by now widely agreed that in cuprates, the mechanism
responsible for superconductivity is closely related to the
unusual antiferromagnetic insulator properties of their normal
states. There are also compelling arguments that the pair
mechanism leading to high-$T\tsub c$ superconductivity does not
correspond to ordinary BCS s-wave pairing. Phase-sensitive
experiments indicate that the SC phase of most cuprates has
$d$-wave-like pairing symmetry (at least for the hole-doped
compounds), and this is also supported by photoemission
experiments, which show the existence of nodal points in the
quasiparticle gap.

Such observations argue strongly for a theory based on continuous
symmetries of the dynamical system that is capable of describing
more sophisticated pairing than found in the simple BCS picture,
and capable of unifying SC and AF collective modes and the
corresponding phases. Then such fundamentally different physics as
SC order and AF order can emerge from the same effective
Hamiltonian as global variables (e.g., doping and temperature) are
varied.

\subsection{Basic ingredients of the SU(4) model}

The basic assumption of the SU(4) model \cite{Gu99} is that the
configuration space is built from {\it coherent pairs} formed from
two electrons (or holes) centered on adjacent lattice sites. [The
SU(4) model is particle--hole symmetric \cite{particle-hole}. For
cuprate superconductors one is generally interested in hole-doped
compounds but more general applications of SU(4) can deal with
either electrons or holes. Hereafter, unless specified explicitly,
we use ``electrons" to reference either electrons or holes.] In
cuprates the coherent pairs are believed to exhibit $d$-wave
orbital symmetry \cite{Sc95}, and we assume a coexistence of two
kinds of coherent pairs in a minimal model: the spin-singlet ($D$)
and the spin-triplet ($\pi$) pairs. We adopt the $d$-wave pairs
with structure defined in Refs.\ \cite{Zh97,De95} as the basic
{\it dynamical building blocks} of the SU(4) model
\begin{equation}
\begin{array}{ll}
\displaystyle D^\dagger\equiv p_{12}^\dagger=\sum_k g(k) c_{k\uparrow}^\dagger
c_{-k\downarrow}^\dagger\ \qquad & D = (D^\dagger)^\dagger,
\\
\displaystyle \pi^\dagger_{ij}\equiv q_{ij}^\dagger = \sum_k g(k)
c_{k+Q,i}^\dagger c_{-k,j}^\dagger \qquad & \pi_{ij} =
(\pi^\dagger_{ij})^\dagger,
\end{array}
\label{D_Pi}
\end{equation}
where  $c_{k,i}^\dagger$ creates an electron of momentum $k$ and
spin projection $i,j= 1 {\rm\ or\ }2 \ (\equiv  \uparrow$ or
$\downarrow)$,
$$
g(k)=(\cos k_x-\sin k_y)
$$
is the $d$-wave form factor, and $Q=(\pi,\pi,\pi)$ is an AF
ordering vector. These pair operators, when supplemented with
operators of particle-hole type ${\cal Q}_{ij}$ and $ S_{ij} $,
\begin{equation}
\begin{array}{c}
 \displaystyle {\cal Q}_{ij} = \sum_k c_{k+Q,i}^\dagger c_{k,j}
\\[1em]
 \displaystyle S_{ij} =
\sum_k c_{k,i}^\dagger c_{k,j} - \tfrac12 \Omega \delta_{ij},
\end{array}
\label{Q_S}
\end{equation}
constitute a 16-element  operator set that is closed under a U(4)
$\supset$ U(1) $\times$ SU(4) algebra if the condition
$$
g(k)\approx\mbox{sgn}\,(\cos k_x-\cos
k_y)
$$
is imposed. In Eq.\ (\ref{Q_S}), $\Omega$ is the maximum number of
doped electrons that can form coherent pairs, assuming the normal
state (at half filling) to be the vacuum. The U(1) factor in U(1)
$\times$ SU(4) is associated with charge-density waves and is
independent of the SU(4) algebra. The charge-density waves can be
excluded in the symmetry limit and in the following discussion we
shall restrict attention to the SU(4) subgroup. The 15 SU(4)
generators are related to more physical operators through the
linear combinations
\begin{eqnarray}
\vec S\ &=& \left( \frac{S_{12}+S_{21}}{2}, \ -i \, \frac
{S_{12}-S_{21}}{2}, \ \frac {S_{11}-S_{22}}{2} \right) \nonumber
\\
\vec{\cal Q}\ &=& \left(\frac{{\cal Q}_{12}+{\cal
Q}_{21}}{2},-i\frac{{\cal Q}_{12}-{\cal Q}_{21}}{2}, \frac{{\cal
Q}_{11}-{\cal Q}_{22}}{2} \right) \nonumber
\\
\vec \pi^\dagger &=& \left(\ i\frac {q_{11}^\dagger\ -
q_{22}^\dagger}2, \ \frac{q_{11}^\dagger + q_{22}^\dagger}2, \
-i\frac {q_{12}^\dagger + q_{21}^\dagger}2 \right)
\label{operatorset}
\\
D^\dagger &=& p^\dagger_{12} \qquad D = p_{12} \nonumber
\\
\hat{n}\ &=&\sum_{k,i}
c_{k,i}^\dagger c_{k,i} =S_{11}+S_{22}+\Omega\ \nonumber
\end{eqnarray}
where $\vec S$ is the spin operator, $\vec{\cal Q}$ the staggered
magnetization, $\vec \pi^\dagger$ ($\vec \pi$) the vector form of
the creation (annihilation) operator of spin-triplet pairs, and
$\hat{n}$  the electron number operator.  It will also sometimes
prove useful to replace the number operator $\hat n$ with the
charge operator $M$, defined through
\begin{equation}
    M=\frac12(S_{11}+S_{22})=\tfrac12 (\hat n-\Omega).
\nonumber
\end{equation}
It has been demonstrated \cite{wu03} that this SU(4) algebra
defines the minimal symmetry implementing a unified description of
antiferromagnetism and $d$-wave superconductivity that is
consistent with Mott-insulator properties for normal states.

\subsection{The group structure and model Hamiltonian}

The SU(4) group has three dynamical symmetry group chains:
\begin{eqnarray}
&\supset& {\rm SO(4)} \times {\rm U(1)}
 \supset {\rm SU(2)} \tsub{s} \times {\rm U(1)} \nonumber \\
{\rm SU(4)} &\supset& {\rm SO(5)} \supset {\rm SU(2)} \tsub{s}
\times {\rm U(1)}
\label{gchain}
\\ &\supset& {\rm SU(2)} \tsub{p}
\times {\rm SU(2)} \tsub{s} \supset {\rm SU(2)}
\tsub{s} \times {\rm U(1)} \nonumber
\end{eqnarray}
each of which ends in the subgroup ${\rm SU(2)}\tsub{s} \times
{\rm U(1)}$ representing total spin and charge conservation. An
SU(4) model Hamiltonian containing one and two-body interactions
can then be determined uniquely,
\begin{eqnarray}
H = \varepsilon\hat{n} &+& v \hat{n}^2 -G_0 D^\dag D
-G_1\vec{\pi}^\dag\cdot\vec{\pi} \nonumber \\
&-& \chi\vec{\cal Q}_\cdot\vec{\cal Q}+\kappa\vec{S}\cdot\vec{S} .
\label{Hsu4}
\end{eqnarray}
In Eq.\ (\ref{Hsu4}), the parameters $G_0$, $G_1$, and $\chi$ are
effective interaction strengths of $d$-wave singlet pairing,
triplet pairing, and staggered magnetization, respectively,
$\varepsilon$ is the average single-electron energy, and $v$ may
be interpreted as the mean value of the two-body interaction. In
later discussions, we shall set $\varepsilon=v=0$ because the
first two terms in Eq.\ (\ref{Hsu4}) are state-independent and
provide only a constant energy. The term $\kappa \vec{S}\cdot
\vec{S}$ will also be ignored for the present discussion,
corresponding to assuming the total spin for the ground state to
be zero.

Using the properties of Lie algebras, we have shown \cite{Gu99}
that analytical solutions for matrix elements can be obtained for
the group chains defined in (\ref{gchain}). Each symmetry limit
has been shown to represent a {\it collective mode} that
corresponds to properties observed in the cuprate phase diagram:
the SO(4) limit is associated with an AF phase occurring when
$\chi$ is dominant ($G_0=G_1<\chi$), the ${\rm SU(2)}\tsub{p}$
limit is associated with a $d$-wave SC phase occurring when $G_0$
is dominant ($\chi=G_1<G_0$), and the SO(5) limit is a critical
dynamical symmetry \cite{lawu99} representing a transitional phase
that is soft against AF and pairing fluctuations (occurring when
$G_0=\chi$). A more extensive discussion of these symmetry limits
and their corresponding phases can be found in Refs.\
\cite{Gu99,lawu99}.

\subsection{Doping dependence in the SU(4) model}

The phenomenology of the cuprate superconductors suggests that the
expectation value of the system Hamiltonian should depend
microscopically on the amount of hole doping. Within the SU(4)
model it is doping that drives the system from one dynamical
symmetry limit to another, causing the system to undergo
crossovers and phase transitions. Let us now explore this
microscopic doping dependence of the SU(4) model in more detail.

The lowest-order Casimir operators $C\tsub g$ for each subgroup g
in (\ref{gchain}) are
\begin{eqnarray}
C\tsub{SO(4)} &=& \vec {\cal Q} \cdot \vec {\cal Q} + \vec S
\cdot\vec
S \nonumber \\
C\tsub{SO(5)} &=&\vec \pi^\dagger \cdot \vec \pi + \vec
S \cdot \vec S +M(M-3) \nonumber \\
C_{\mbox{\scriptsize SU(2)}_{\mbox{\scriptsize p}}} &=&D^\dagger
\cdot D +M(M-1)
\nonumber \\
C_{\mbox{\scriptsize SU(2)}_{\mbox{\scriptsize s}}} &=& \vec S
\cdot \vec S \nonumber
\\
C\tsub{U(1)} &=&M \mbox{ and }M^2.
\nonumber
\end{eqnarray}
The SU(4) quadratic Casimir operator is
\begin{equation}
C\tsub{SU(4)} =D^\dagger \cdot D +\vec \pi^\dagger \cdot \vec \pi
+\vec {\cal Q} \cdot \vec {\cal Q} + \vec S \cdot\vec S +M(M-4),
\label{su4casimir}
\end{equation}
which is an invariant in the SU(4) representation space.

In the general case we may introduce a seniority-like quantum
number $n_b=u\Omega$, which is the number of particles in the
system that do not couple to $D$ or $\pi$ pairs, with $u$ being
the number density of the unpaired particles. For the $u=0$ case,
we take as a Hilbert space
\begin{eqnarray}
\ket{u=0} &=& \ket{n_x n_y n_z n_d}
\nonumber
\\
&=& (\pi_x^\dagger)^{n_x} (\pi_y^\dagger)^{n_y}
(\pi_z^\dagger)^{n_z} (D^\dagger)^{n_d}
\ket{0},
\nonumber
\end{eqnarray}
which is a collective subspace (the $D$--$\pi$ pair space)
associated with SO(6) irreps of the form $
(\sigma_1,\sigma_2,\sigma_3) = \left(\tfrac \Omega2,0,0\right)$.
[Because SO(6) and SU(4) have the same Lie algebra we choose to
label SU(4) irreps with SO(6) quantum numbers.] The expectation
value of the SU(4) Casmir operator (\ref{su4casimir}) for $u=0$ is
\begin{equation}
\langle C\tsub{SU(4)}\rangle =\frac\Omega2
\left(\frac\Omega2+4\right),
\nonumber
\end{equation}
and is a constant for any state in the $D$--$\pi$ pair space. More
generally, states with $n_b$ unpaired particles can be described
by irreps of the form $\left(\tfrac12 (\Omega-n_b), 0, 0\right)$.
Thus for the $u\ne 0$ case with broken pairs the expectation value
of the SU(4) Casimir operator is
\begin{equation}
\bra{u} C\tsub{SU(4)}\ket{u}=\frac\Omega2
\left(\frac\Omega2+4\right) - \frac{n_b}{2}
\left(\Omega-\frac{n_b}{2}+4\right).
\nonumber
\end{equation}

For a system with charge $\langle M\rangle=\frac12(n-\Omega)$, the
SU(4) invariance leads to a conserved quantity
\begin{eqnarray}
{\cal E}\tsub{SU4}&=&D^\dag D
             +\vec{\pi}^\dag\cdot\vec{\pi}
             +\vec{\cal Q}\cdot\vec{\cal Q}+\vec{S}\cdot\vec{S}
\nonumber\\
&=&{C}\tsub{SU4}-M(M-4). \nonumber
\end{eqnarray}
The expectation value $\langle{\cal E}\tsub{SU4}\rangle$ is thus a
constant, independent of how the system changes its state within
the SU(4) space.  It follows that in the large $\Omega$ limit,
\begin{equation}
\langle {\cal E}\tsub{SU4}\rangle= \frac
{\Omega^2}4\left[(1-u)^2-x^2\right] \qquad x=1-\frac n\Omega\ ,
\label{su4const}
\end{equation}
where $n$ is the electron number. In the above expressions, $x$
may be regarded as the relative doping fraction in our theory.
Since $\Omega-n$ is the hole number when $n<\Omega$, positive $x$
represents the case of hole-doping, with $x=0$ corresponding to
half filling (no doping) and $x=1$ to maximal hole-doping.
Negative $x$ ($n>\Omega$) is defined naturally as the relative
doping fraction for electron-doping.

The true doping rate, defined as $P = (\Omega-n)/\Omega_e$, where
$\Omega_e$ is the number of lattice sites, is related to $x$
through
\begin{equation}
P=x\frac\Omega{\Omega_e}=xP_{\mbox{\scriptsize f}},
\qquad
P_{\mbox{\scriptsize f}}\equiv\frac\Omega{\Omega_e},
\nonumber
\end{equation}
with $P>0$ for hole doping and $P <0$ for electron doping.
$P_{\mbox{\scriptsize f}}$ can be regarded as the maximum possible
value of the true doping rate $P$. Experimentally,
$P_{\mbox{\scriptsize f}}$ is found to be 0.23 $\sim$ 0.27
\cite{Ti99}.

\subsection{No-double-occupancy and maximum doping}

The SU(4) group is the minimal symmetry accommodating
superconductivity and antiferromagnetism in cuprates. It is
further found that the SU(4) symmetry is a consequence of
non-double-occupancy -- the constraint that each lattice site
cannot have more than one valence electron. It has been
demonstrated \cite{wu03} that with no approximation to the
$d$-wave formfactor $g(k)$, the momentum-space operator sets
(\ref{D_Pi}) and (\ref{Q_S}) may be expressed in the coordinate
space as
\begin{eqnarray}
p_{12}^\dagger&=&\sum_{r= {\rm even}} \left( c_{{\bf
r}\uparrow}^\dagger c^\dagger_{\bar{\bf r}\downarrow}\ -\ c_{{\bf
r}\downarrow}^\dagger c^\dagger_{\bar{\bf r}\uparrow}\right) \nonumber \\
q_{ij}^\dagger&=&\sum_{r= {\rm even}} \left( c_{{\bf
r},i}^\dagger c^\dagger_{\bar{\bf r},j}+c_{{\bf r},j}^\dagger
c^\dagger_{\bar{\bf r},i}\right) \nonumber \\
S_{ij}&=&\sum_{r= {\rm even}} \left( c_{{\bf r},i}^\dagger
c_{{\bf r},j}-c_{\bar{\bf r},j} c_{\bar{\bf r},i}^\dagger\right)
\label{rspace2}\\
\tilde{Q}_{ij}&=&\sum_{r= {\rm even}} \left( c_{{\bf r},i}^\dagger c_{{\bf
r},j}+c_{\bar{\bf r},j}c_{\bar{\bf r},i}^\dagger\right)
\nonumber
\\
p_{12}&=&(p_{12}^\dagger)^\dagger \qquad
q_{ij}=(q_{ij}^\dagger)^\dagger .
\nonumber
\end{eqnarray}
In Eq.\ (\ref{rspace2}), the summation is over even sites of the
lattice, the quantity $\tilde{Q}_{ij}$ is defined by
$$
\tilde{Q}_{ij}\equiv {\cal Q}_{ij}+ \tfrac12 \delta_{ij}\Omega,
$$
and $c^\dagger_{{\bf r},i}$ ($c_{{\bf r},i}$) creates
(annihilates) an electron of spin $i$ located at ${\bf r}$, while
$c^\dagger_{\bar{{\bf r}},i}$ ($c_{\bar{{\bf r}},i}$) creates
(annihilates) an electron of spin $i$ at its four neighboring
sites, ${\bf r}\pm{\bf a}$ and ${\bf r}\pm{\bf b}$, with equal
probabilities (${\bf a}$ and ${\bf b}$ are the crystal constants
along the ${\bf x}$ and ${\bf y}$ directions, respectively,
 on the copper-oxide plane),
\begin{equation}
c^\dagger_{\bar{\bf r},i}=\frac 12 \left(c^\dagger_{{\bf r} +{\bf
a},i}+c^\dagger_{{\bf r}-{\bf a},i} -c^\dagger_{{\bf r}+{\bf b},i}
-c^\dagger_{{\bf r}-{\bf b},i}\right) .
\nonumber
\end{equation}
Explicit commutation operations show that only if the
no-double-occupancy constraint is imposed, so that the
anticommutator relation
\begin{equation}
\left\{c_{\bar{\bf r},i}\ , c^\dagger_{\bar{\bf
r}',i'}\right\}=\delta_{\bf rr'}\delta_{ii'}
\nonumber
\end{equation}
is valid, does the operator set (\ref{rspace2}) close under the
${\rm U(1)} \times {\rm SU(4)}$ Lie algebra. Thus, the
coordinate-space commutation algebra of the operators
(\ref{rspace2}) demonstrates that SU(4) symmetry {\em necessarily
implies} a no-double-occupancy constraint in the copper oxide
conducting plane. This suggests a fundamental relationship between
SU(4) symmetry and Mott-insulator normal states at half filling
for cuprate superconductors.

One immediate conclusion \cite{wu03} is that the
no-double-occupancy constraint sets an upper limit for the number
of doped holes if SU(4) is to be preserved exactly:
$\Omega\tsub{max} = \tfrac14 \Omega_e$. Thus, the maximum doping
fraction consistent with SU(4) symmetry is
\begin{equation}
P_{\mbox{\scriptsize f}}\equiv\frac\Omega{\Omega_e}=\frac 1 4 .
\nonumber
\end{equation}
Beyond this doping fraction the no-double-occupancy condition
cannot be ensured and exact SU(4) symmetry cannot be preserved.
The empirical maximum doping fraction 0.23 $\sim$ 0.27 for cuprate
superconductivity may then be taken as indirect evidence for a
strongly-realized SU(4) symmetry underlying the superconductivity
in cuprates.

\section{Gap Equations in the SU(4) Model}

Equation (\ref{gchain}) defines three dynamical symmetry group
chains, each representing a collective mode. Analytical solutions
associated with these symmetry limits have been derived
\cite{Gu99}. However, a realistic physical system may not lie in
any of these dynamical symmetry limits. In order to study
properties of a realistic system and phase transitions between the
symmetry limits, one needs systematic ways to deal with
approximate symmetries. There is a well-developed theoretical
approach to relating a many-body algebraic theory to an
approximation of that theory that exhibits spontaneous symmetry
breaking: the method of generalized coherent states
\cite{Zh90,lawu99}. This method is a variational procedure using
the coherent state as the trial wavefunction in a quasiparticle
space. It may be viewed as the most general
Hartree--Fock--Bogoliubov variational formulation, but with an
additional proviso that the variational states are constrained to
respect the highest symmetry of a set of group chains, as in Eq.\
(\ref{gchain}). Generalized coherent states are thus particularly
suitable for studying the ground state properties of any strongly
correlated many-body system that is amenable to a dynamical
symmetry description.

\subsection{The generalized coherent-state method}

The coherent state $|\psi\rangle$ associated with the SU(4)
symmetry can be written formally as
\begin{equation}
|\psi\rangle={\cal T}\mid 0^* \rangle,
\label{eq10}
\end{equation}
with the operator ${\cal T}$ defined by
\begin{equation}
{\cal T} = \exp (\eta_{00} p_{12}^{\dagger }+\eta_{10}
q_{12}^{\dagger }-{\rm h.\ c.}).
\label{eq10b}
\end{equation}
In Eq.\ (\ref{eq10}), $|0^{*}\rangle$ is the physical vacuum (the
ground state of the system), the real parameters $\eta_{00}$ and
$\eta_{10}$ are symmetry-constrained variational parameters, and
h.\ c.\ stands for the Hermitian conjugate. Since the variational
parameters weight the elementary excitation operators
$p^\dagger_{12}$ and $q^\dagger_{12}$ in Eq.\ (\ref{eq10b}), they
represent collective state parameters for a $D$-$\pi$ pair
subspace truncated under the SU(4) symmetry
\cite{coherent-parameters}.

The symmetry-constrained variational Hamiltonian is
\begin{equation}
H' = H-\lambda\hat{n},
\nonumber
\end{equation}
where $H$ is the Hamiltonian (\ref{Hsu4}) and $\lambda$ is the
chemical potential, determined by requiring particle-number
conservation. The parameters $\eta_{00}$ and $\eta_{10}$ in Eq.\
(\ref{eq10b}) are determined by the variational principle
$\delta\langle H'\rangle=0$, where $\langle H'\rangle$ is the
expectation value of $H'$ with respect to the ground state
$|0^*\rangle$
$$
\langle H'\rangle\equiv\langle 0^*| H'|0^*\rangle.
$$

As shown in Appendix A, it is convenient to evaluate the variation
$\delta\langle H'\rangle=0$ using a 4-dimensional matrix
representation that was introduced in Refs.\
\cite{lawu99,wmzha88}. In this representation the unitary operator
${\cal T}$ implements a transformation from the original particle
basis to a quasiparticle basis and the variation parameters
$\eta_{00}$ and $\eta_{10}$ are replaced, respectively, by $u_\pm$
and $v_\pm$, with a unitary condition $u^2_\pm+v^2_\pm=1$. Under
this transformation the basic fermion operators
$$
\{c^\dagger_{{\bf r}\uparrow},c^\dagger_{{\bf r}\downarrow},
c_{\bar{\bf r}\uparrow},c_{\bar{\bf r}\downarrow} \}
$$
are converted to quasiparticle operators
$$
\{a^\dagger_{{\bf r}\uparrow}, a^\dagger_{{\bf r}\downarrow},
a_{\bar{{\bf r}}\uparrow}, a_{\bar{{\bf r}}\downarrow} \} ,
$$
as shown explicitly in Eq.\ (\ref{transfA}). This implies that
\begin{eqnarray}
\left( u_+ c_{{\bf r}\uparrow} + v_+ c^\dagger_{{\bar{\bf
r}}\downarrow} \right) |0^*\rangle &=& a_{{\bf r}\uparrow}
|\psi\rangle
\nonumber\\
\left( u_- c_{{\bf r}\downarrow} - v_- c^\dagger_{{\bar{\bf
r}}\uparrow} \right) |0^*\rangle &=& a_{{\bf r}\downarrow}
|\psi\rangle
\nonumber\\
\left( u_- c^\dagger_{{\bar{\bf r}}\uparrow} + v_- c_{{\bf
r}\downarrow} \right) |0^*\rangle &=& a^\dagger_{{\bar{\bf
r}}\uparrow} |\psi\rangle
\nonumber\\
\left( u_+ c^\dagger_{{\bar{\bf r}}\downarrow} - v_+ c_{{\bf
r}\uparrow} \right) |0^*\rangle &=& a^\dagger_{{\bar{\bf
r}}\downarrow} |\psi\rangle
\nonumber
\end{eqnarray}
and one sees that this is a Bogoliubov-type transformation: each
quasiparticle state is a mixture of a particle and a hole, and the
coherent state $|\psi\rangle$ is an SU(4)-symmetry constrained
quasiparticle vacuum.

By using the matrix representation one can calculate expectation
value for any operator $\hat O$ in the coherent state
representation through the transformation
$$
\langle 0^*|\hat
O|0^*\rangle = \langle\psi |{\cal T}\hat O{\cal
T}^{-1}|\psi\rangle.
$$
Detailed derivations are given in Appendix A.

\subsection{Temperature dependence}

At finite temperature, $|\psi\rangle$ may no longer be a
quasiparticle vacuum state and the quasiparticle annihilation
operators acting on $|\psi\rangle$ do not necessarily give zero.
In Appendix B a formalism is derived to deal with the
finite-temperature case. To formulate the simplest initial model
we assume that at a temperature $T$ the single-particle levels
$\varepsilon_{r\pm}$ (defined in Eq.\ (\ref{qse}) below) are
degenerate and contain $\tilde{n}_{r+} +\tilde{n}_{r-}$
quasiparticles. The quasiparticle number densities are then
assumed to be given by the Fermi-Dirac distribution
\begin{equation}
\tilde{n}_{\pm} (T) =\frac
2{\Omega}\sum_{r=even}\tilde{n}_{r\pm}(T) =\frac{2}{1+\exp (R
e_{\pm}/k\tsub B T)} ,
\label{thermal}
\end{equation}
where $e_\pm$ is the quasiparticle energy defined in Eq.\
(\ref{qpe}) below. In Eq.\ (\ref{thermal}), we have introduced an
energy scaling factor $0<R<1$. This is because the actual
single-particle energies are generally non-degenerate so that the
realistic quasiparticle excitation should be easier. Thus
$\tilde{n}_{\pm}(T)$ should in general be larger than those with
the degenerate approximation. The factor $R$ accounts for this
effect in an average manner, and it may be determined by fitting
to data.

For one-body operators at finite temperatures the expectation
values are (see Appendix B)
\begin{eqnarray}
\langle D^{\dagger}\rangle &=&\langle D\rangle=-\tfrac{\Omega}{2}
\left[P_+(T)u_+v_+ + P_-(T)u_-v_-\right]
\nonumber
\\
\langle \pi^{\dagger}_z\rangle &=&\langle \pi_z\rangle=-\tfrac
{\Omega}{2} \left[P_+(T)u_+v_+ -P_-(T)u_-v_-\right]
\nonumber
\\
\langle {\cal Q}_z\rangle &=& \tfrac{\Omega}{2} \left[ P_+(T)v_+^2
- P_-(T)v_-^2\right]
\label{eqone2}
\\
\langle\hat n\rangle &=& \tfrac{\Omega}{2} \left[P_+(T)(2v_+^2-1)
+P_-(T) (2v_-^2-1)+2\right] \nonumber
\\
\langle \pi_x\rangle &=& \langle \pi_y\rangle=\langle
\vec{S}\rangle = \langle {\cal Q}_x\rangle=\langle {\cal
Q}_y\rangle=0 ,
\nonumber
\end{eqnarray}
where we have defined
\begin{equation}
P_{\pm}(T)= 1-\tilde{n}_{\pm}(T) =\tanh \left(\frac{R
e_{\pm}}{2k\tsub B T} \right).
\nonumber
\end{equation}
For the scalar products of these one-body operators the
expectation values are products of the corresponding one-body ones
in the large-$\Omega$ approximation,
\begin{eqnarray}
\langle D^{\dagger }D\rangle &=& \langle D\rangle^2
\nonumber
\label{eqtwo2}
\\
\langle \vec{\pi}^{\dagger
}\cdot\vec{\pi}\rangle &=& \langle \pi_z\rangle^2
\\
\langle \vec{\cal Q} \cdot \vec{\cal Q}\rangle &=& \langle {\cal Q}_z\rangle^2
\nonumber
\\
\langle \vec{S} \cdot \vec{S}\rangle &=& 0 .
\nonumber
\end{eqnarray}
If $T\rightarrow 0$, then $P_\pm(T)\rightarrow 1$ and Eqs.\
(\ref{eqone2}) and (\ref{eqtwo2}) reduce to Eqs.\ (\ref{eqoneA})
and (\ref{eqtwoA}), respectively.

\subsection{Energy gaps and gap equations}

We now use the preceding results to express the variational
Hamiltonian $\langle H'\rangle$ in the coherent state
representation. Introducing the energy gaps
\begin{eqnarray}
\Deltad&\equiv&G_0\sqrt{\left\langle D^\dag D\right\rangle}
\nonumber\\
\Deltapi&\equiv&G_1\sqrt{\left\langle\vec{\pi}^\dag\cdot
            \vec{\pi}\right\rangle}
\label{1.11}\\
\Deltaq&\equiv&\chi\sqrt{\left\langle\vec{\cal Q}\cdot
            \vec{\cal Q}\right\rangle} ,
\nonumber
\end{eqnarray}
one obtains
\begin{equation}
\langle  H' \rangle = (\varepsilon-\lambda) n -\left (\
\frac{\Deltadsq}{G_0} +\frac{\Deltapisq}{ G_1}+\frac{\Deltaqsq}{
\chi}\ \right).
\nonumber
\end{equation}
Variation of $\langle H'\rangle$ with respect to $u_\pm$ or
$v_\pm$ (that is, solving $\delta \langle H'\rangle=0$)
yields
\begin{equation}
2u_{\pm}v_{\pm} (\varepsilon_{ \pm}-\lambda)-\Deltapm(u^2_{\pm}
-v^2_{\pm})=0,
\nonumber
\end{equation}
which is satisfied by
\begin{equation}
\begin{array}{c}
\displaystyle u^2_{\pm} = \frac{1}{2}\left
(1+\frac{\varepsilon_{\pm}-\lambda}{e_{\pm}}\right)
\\[1.5em]
\displaystyle v^2_{\pm} = \frac{1}{2}\left
(1-\frac{\varepsilon_{\pm}-\lambda}{e_{\pm}} \right), \label{uv}
\end{array}
\end{equation}
where
\begin{equation}
e_{\pm} = \sqrt{(\varepsilon_{\pm}-\lambda)^2+{\Deltapm}^2}
\label{qpe}
\end{equation}
and
\begin{equation}
\Deltapm = \Deltad\pm\Deltapi \qquad
\varepsilon_{\pm}=\varepsilon\mp\Deltaq .
\label{qse}
\end{equation}
Inserting Eq.\ (\ref{uv}) into Eqs.\ (\ref{eqone2} --
\ref{eqtwo2}) and employing the gap definitions (\ref{1.11}), one
obtains the temperature-dependent gap equations
\begin{subequations}
\label{gap:whole}
\begin{eqnarray}
\Deltad&=&\frac{G_0\Omega}{4}\left ( w_+ \DeltaPlus  + w_-
\DeltaMinus\right )
\label{subgapeq:d}\\
\Deltapi&=&\frac{G_1\Omega}{4} \left (  w_+ \DeltaPlus - w_-\
\DeltaMinus \right )
\label{subgapeq:pi}\\
\frac{4\Deltaq}{\chi\Omega}&=& w_+ ( \Deltaq+\lambdaPrime ) + w_-
( \Deltaq-\lambdaPrime )
\label{subgapeq:q}\\
-2x&=&w_+ ( \Deltaq+\lambdaPrime ) - w_- ( \Deltaq-\lambdaPrime ),
\label{subgapeq:la}
\end{eqnarray}
\end{subequations}
where we define
\begin{equation}
w_\pm \equiv \frac {P_\pm(T)}{e_\pm} ,
\label{wpm}
\end{equation}
and
\begin{equation}
\lambdaPrime \equiv \lambda-\varepsilon\ .
\label{1.22}
\end{equation}
By solving the above algebraic equations, all the gaps and the
chemical potential $\lambdaPrime$ can be obtained. The total
energy can be calculated as
\begin{eqnarray}
E&=&\left \langle H'\right\rangle+\lambda n \nonumber
\\
&=&\varepsilon n -\left(\frac{\Deltadsq}{G_0}
        +\frac{\Deltapisq}{G_1}+\frac{\Deltaqsq}{\chi} \right).
\nonumber
\end{eqnarray}
To simplify the discussion, we shall hereafter ignore the
single-particle energy in the above equation by setting
$\varepsilon=0$, since this term has been approximated as a
state-independent constant in our model and thus plays no role in
the phase competition. We can then express the energy density
$E/\Omega$ as
\begin{equation}
 \frac{E}{\Omega}=
-\left ( \frac{{\Deltad}^2}{G_0\Omega
}+\frac{{\Deltapi}^2}{G_1\Omega } +\frac{{\Deltaq}^2}{\chi\Omega }
\right ) .
\label{eqED}
\end{equation}

The three gaps $\Deltad$, $\Deltapi$ and $\Deltaq$ in the above
equations represent, respectively, the characteristic energy
scales of spin-singlet pairing, triplet pairing, and the AF
correlation. Hence the ground state energy is determined by the
three energy gaps. Once the gaps and the chemical potential
$\lambdaPrime$ are known, the quasiparticle energies $e_\pm$, as
well as the amplitudes $u_\pm$ and $v_\pm$, can all be determined
through Eqs.\ (\ref{uv} -- \ref{qse}), permitting other ground
state properties to be calculated.

These results are formally analogous to those of the BCS theory
with $v_{\pm}^2$ the probability of single particle levels
$\varepsilon_{\pm}$ being occupied, $\Deltapm$ the energy gaps,
and $e_{\pm}$ the quasiparticle energies.  The essential
difference from BCS theory is that conventional pairing theories
deal with one pairing gap and one kind of quasiparticle; here we
have two kinds of quasiparticles and several energy gaps, implying
a large variety of new physics.

In the formalism describing this more sophisticated pairing the
quantities $e_{\pm}$ are energies for two kinds of quasiparticle
excitation, corresponding to two sets of non-degenerate single
particle energy spectra $\{\varepsilon_{\pm}\}$ separated by an
energy $2\Deltaq$. Each level can be occupied by only one electron
of either up or down spin. The corresponding pairing gaps are
$\Deltapm$, which are linear combinations of the two gaps
$\Deltad$ and $\Deltapi$. The probabilities for single-particle
levels to be occupied or unoccupied are $v^2_{\pm}$ and
$u^2_{\pm}$, respectively.

In the following sections we shall give analytical solutions for
the gap equations (\ref{gap:whole}), first for zero temperature
and then for finite temperatures. As we shall see, a rich phase
structure emerges naturally in these solutions as a consequence of
competition between the various energy scales.

\section{Solution of Gap Equations and the Gap Diagram at T = 0}

There are three parameters, $\chi$, $G_0$, and $G_1$, in the
coupled algebraic equations (\ref{gap:whole}), corresponding to
the three elementary interactions in the SU(4) model: the AF
correlation ($\chi$), the spin-singlet pairing ($G_0$), and the
spin-triplet pairing ($G_1$). Physical solutions of the gap
equations depend on the choices for these parameters. Experimental
evidence suggests that these three interactions in cuprates are
all attractive, and we will demonstrate later that the AF
correlation should be the strongest and the spin-triplet pairing
the weakest. Thus, in the results presented below we assume
\begin{equation}
\chi > G_0 > G_1 > 0 .
\label{condition}
\end{equation}
Analytical solutions for the gap equations assuming this condition
can be obtained as follows.

\subsection{Solution of the gap equations at T=0}

The gap solutions at $T=0$ can be written explicitly for two
doping regimes separated by a special doping point given by
\begin{equation}
x_q=\sqrt{\frac{\chi-G_0}{\chi-G_1}}.
\label{xq}
\end{equation}
We shall interpret $x_q$ as a {\em critical doping} marking a
quantum phase transition because the wavefunctions and physical
properties of the two doping regions lying on either size of this
point at zero temperature are qualitatively different.
Specifically, one finds the following solutions. (The general
derivations of the gap solutions are given in Appendix C.)

(1) The all-gap solution for $x\le x_q$:
\begin{subequations}
\label{gapT0:whole}
\begin{eqnarray}
\Deltaq&=&\frac{\chi\Omega}{2}\sqrt{(x^{-1}_q- x)(x_q-x)}
\label{udopDq}
\\
\Deltad&=&\frac{G_0\Omega}{2}\sqrt{x (x_q^{-1}- x)}
\label{udopDd}
\\
\Delta_\pi&=&\frac{G_1\Omega}{2}\sqrt{x (x_q- x)}
\label{udoppi}
\\
\lambdaPrime &=&-\frac{\chi\Omega}{2} x_q (1-x_q
x)-\frac{G_1\Omega}2x.
\label{udopL}
\end{eqnarray}
\end{subequations}
This is the most important solution that has all gaps non-zero and
exists only for the doping range $x\le x_q$. In addition, there
are trivial solutions in which at least two gaps are zero.

(2) The pure spin-singlet pairing solution with
$\Delta_q=\Delta_\pi=0$ is valid for the entire physical doping
range $0\le x\leq 1$ and given by
\begin{subequations}
\label{gapDd:whole}
\begin{eqnarray}
\Deltaq&=&\Delta_\pi= 0
\label{odopDq}\\
\DeltaZero &\equiv& \Deltad =\frac{G_0\Omega}{2}\sqrt{1- x^2}
\label{odopDd}\\
\lambdaPrime\hspace{3pt} &=&-\frac{G_0\Omega}{2}x.
\label{odopL}
\end{eqnarray}
\end{subequations}
This is also an important solution in that it gives the
ground-state for $x>x_q$ (see discussions below). It can be
verified easily that both solutions (\ref{gapT0:whole}) and
(\ref{gapDd:whole}) satisfy the SU(4) condition (\ref{su4const})
with $u=0$, which means that all electrons are paired at $T=0$.

For completeness, we list below additional trivial solutions.
These are a spin-triplet pairing solution, a pure AF solution, and
a metal solution (with all gaps zero), all of which are valid for
the entire physical doping range $0\leq x\leq 1$:

(3) The spin-triplet pairing solution:
\begin{subequations}
\label{gapDpi:whole}
\begin{eqnarray}
\Deltaq&=&\Deltad= 0
\label{eqDqd}\\
\Deltapi &=&\frac{G_1\Omega}{2}\sqrt{1- x^2}
\label{eqDpi}\\
\lambdaPrime &=&-\frac{G_1\Omega}{2}x.
\label{eqLpi}
\end{eqnarray}
\end{subequations}

(4) The pure AF solution:
\begin{subequations}
\label{gapDaf:whole}
\begin{eqnarray}
\Deltad&=&\Deltapi= 0
\label{eqDdpi}\\\
\Deltaq &=&\frac{\chi\Omega}{2}(1- x)
\label{eqdaf}\\
\lambdaPrime &=&-\frac{\chi\Omega}{2}(1-x).
\label{eqLaf}\
\end{eqnarray}
\end{subequations}

(5) The metal solution:
\begin{subequations}
\label{gapMetal:whole}
\begin{eqnarray}
\Deltaq&=&\Deltad=\Deltapi=0,
\label{gap0metal} \\
\lambdaPrime&=&-2kT\ \atanh(x)
\label{metal}\\
&=&0\ \mbox{ at $T=0$} .
\nonumber
\end{eqnarray}
\end{subequations}

The trivial solutions can be verified easily. For example, when
$\Delta_q=0$ one obtains immediately $w_+=w_-$ from Eq.\
(\ref{subgapeq:q}). There are two solutions satisfying Eqs.\
(\ref{subgapeq:d}) and (\ref{subgapeq:pi}): either $\Deltad\ne 0$
and $\Deltapi= 0$, or $\Deltapi\ne 0$ and $\Deltad= 0$.  The
former gives the singlet pairing solution (\ref{odopDq} --
\ref{odopDd}) while the latter leads to the triplet pairing
solution (\ref{eqDqd} -- \ref{eqDpi}). Eqs.\ (\ref{odopL}) and
(\ref{eqLpi}) for $\lambdaPrime$ result from Eq.\
(\ref{subgapeq:la}). The AF solution is obtained because
$\Deltad=\Deltapi= 0$ is a trivial solution of Eqs.\
(\ref{subgapeq:d} -- \ref{subgapeq:pi}). The metallic state
solution with all gaps vanishing is obvious from inspection of the
equations.

\epsfigbox{fg x}{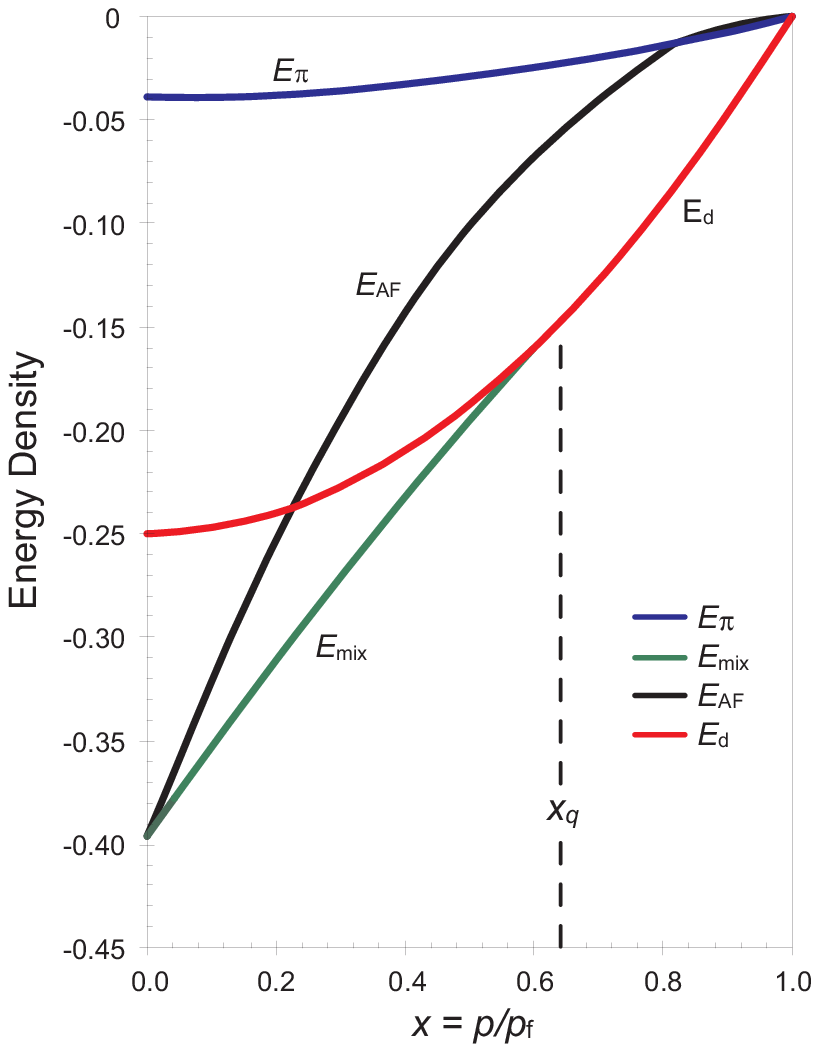}{0pt}{0pt} {(Color online) Total
energy associated with different gap solutions at $T=0$.
$E\tsub{mix}$ is the energy calculated with the all-gap solution
(\ref{gapT0:whole}), while $E\tsub{d}$ (calculated from the
solution in Eqs.\ (\ref{gapDd:whole})), $E_\pi$ (Eqs.\
(\ref{gapDpi:whole})) and $E\tsub{AF}$ (Eqs.\
(\ref{gapDaf:whole})) represent, respectively, the energy density
of the spin-singlet pairing, the spin-triplet pairing, and the AF
solutions. The energy of the metallic solution is set to be zero
and taken as the energy reference. The interaction strengths used
in this plot are the same as those in Fig.\ \ref{fig:fg 1}. }

We shall demonstrate below that these solutions contain rather
rich physics as a function of doping. Among the five sets of gap
solutions (Eqs.\ (\ref{gapT0:whole} -- \ref{gapMetal:whole})), the
one with the lowest energy at each doping corresponds to the
physical ground state. We can calculate these energies by
inserting the gap solutions directly into Eq.\ (\ref{eqED}), and
then investigate how these different sets of solutions compete
with each other at $T=0$. From Fig.\ \ref{fig:fg x}, one sees
clearly that the energy of the all-gap solution $E\tsub{mix}$ is
always the lowest one and thus is the physical ground state for
the doping range $x\leq x_q$. For $x>x_q$, the pure singlet
pairing state becomes the ground state because $E\tsub{d}$ is the
lowest energy for this doping range. All the other trivial
solutions lie higher in energy. They may be regarded as collective
excited states but they cannot become the physical ground state at
$T=0$.

As we will see later in the study of phase transitions, for $T>0$
the AF or the metallic state could become the ground state in
certain temperature and doping ranges. However, note that this can
never happen for the spin-triplet pairing state, as long as $G_1$
is the weakest of the three coupling parameters. In other words,
although spin-triplet pairing plays an important role in the SU(4)
theory, the pure spin-triplet state can never be reached by
thermal excitations as long as the condition (\ref{condition}) is
satisfied.

\subsection{Gap diagram at T=0}

\epsfigbox{fg 1}{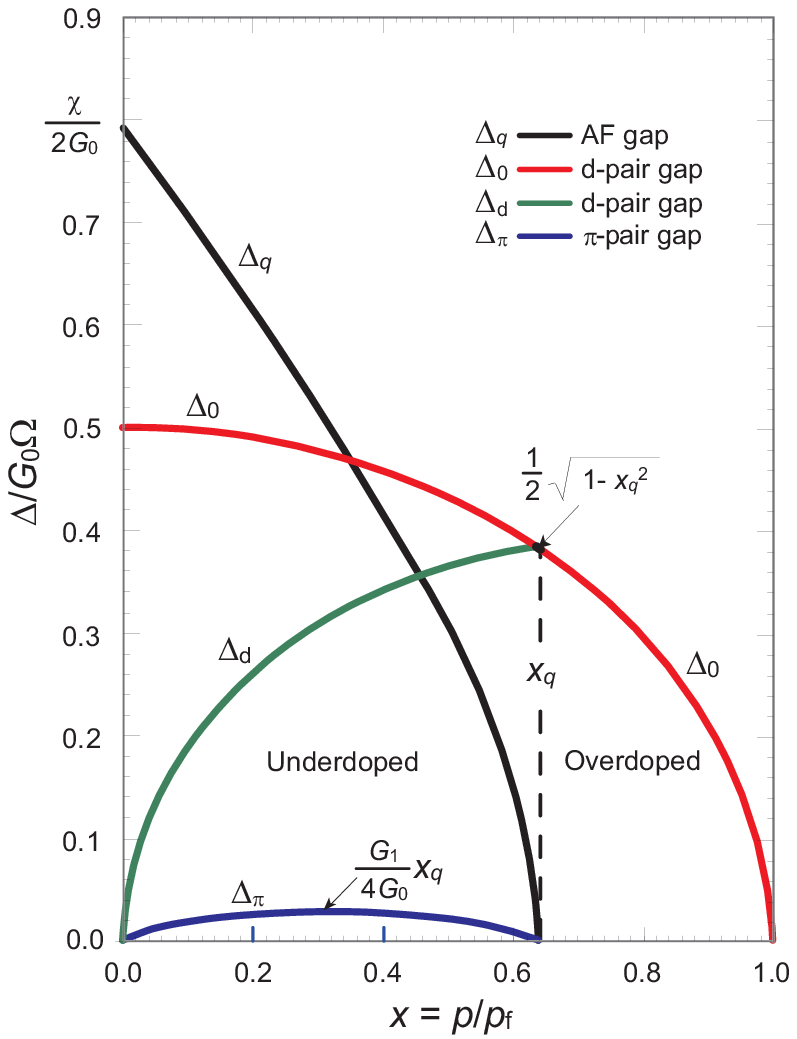}{0pt}{0pt} {(Color online) A generic gap
diagram for energy gaps versus doping, as predicted by the SU(4)
model at $T=0$. The energy gaps are scaled by $G_0\Omega$ and the
doping parameter is scaled by the maximum doping $P\tsub f$ (We
assume $P\tsub f=1/4$ \protect\cite{wu03}). The interaction
strengths are assumed to be $\chi = 13$, $G_0 = 8.2$, and $G_1 =
1.3$ (in an arbitrary energy unit), which, according to Eq.\
(\ref{xq}), requires the critical doping point to be $x_q=0.64$. }

A generic gap diagram at $T=0$ describing features of the energy
gaps as functions of doping $x$ is shown in Fig.\ \ref{fig:fg 1}.
The diagram is constructed using the analytical expressions
(\ref{gapT0:whole}) and (\ref{gapDd:whole}). It is {\it generic}
because the relative sizes of the gaps can be modified by the
choice of interaction strengths but their basic forms are dictated
entirely by the algebraic structure of the model [and that, in
turn, is determined by the physically-motivated choice of
generators given in Eq.\ (\ref{operatorset})]. Four
doping-dependent energy gaps are predicted:
\begin{enumerate}

\item The gap $\Deltaq$ (Eq.\ (\ref{udopDq})) measuring
antiferromagnetic correlations; $\Deltaq$ has its maximal value at
$x=0$, decreases nearly linearly to the region of the pairing gaps
as doping increases, crosses the pairing gaps, and vanishes
eventually at the critical doping $x_q$.

\item The spin-singlet pairing gap $\Deltad$ (Eq.\
(\ref{udopDd})), which is the superconducting gap for $x<x_q$.

\item The spin-singlet pairing gap $\DeltaZero= \Deltad$ (Eq.\
(\ref{odopDd})), which is the superconducting gap for $x>x_q$ but
is not the order parameter for the ground state in the doping
range $x\leq x_q$.

\item The spin-triplet pairing gap $\Deltapi$ (Eq.\
(\ref{udoppi})). Similar to the case for $\Deltaq$, the triplet
gap $\Deltapi$ exists only in the doping range $x \le x_q$. It has
its maximal value at $x_q/2$ and vanishes at both ends of its
range.

\end{enumerate}
Note that the spin-singlet pairing gap exhibits fundamentally
different behavior on the left and right sides of $x_q$: for $x >
x_q$, it corresponds to a monotonic curve, labelled as
$\DeltaZero$. However, as the doping decreases from $x_q$ the
spin-singlet gap splits into two curves (labelled as $\Deltad$ and
$\DeltaZero$) having very different doping dependence. Moreover,
the splitting correlates strongly with the amount of hole doping:
the lower the hole doping, the larger the splitting.

The qualitative features of the SU(4) gap diagram seem to agree
with a large body of recent observations in cuprates
\cite{Shen03}. In particular, the appearance of a critical doping
point and the splitting of the pairing gap in the underdoped
compounds are basic predictions of the SU(4) model that have some
experimental support \cite{Ta01,Ta03}.

\subsubsection{Critical doping point and pairing gap splitting}

The occurrence of a critical doping point and the splitting of the
SC pairing gap are understood in the SU(4) model as a direct
consequence of competing SC pairing and AF correlation in the
doping range below $x_q$.  When doping is small the AF correlation
dominates the SC pairing. In this case, a state with large AF
correlations and suppressed pairing is favored in energy, and thus
can become the ground state. Therefore, the SC gap $\Deltad$ for
the ground state is smaller and the larger pairing gap
$\DeltaZero$ is associated with an excited state. However, as
doping increases the pairing correlation grows quickly and
eventually dominates. The critical doping point $x_q$ is just the
doping fraction at which the AF correlation is completely
suppressed.

The critical doping point $x_q$ defines a natural boundary between
{\em overdoped} and {\em underdoped} regimes having qualitatively
different wavefunctions. It corresponds to the doping point where
AF correlations vanish and separates a doping regime characterized
by weak superconductivity and reduced pair condensation energy
from a doping regime characterized by strong superconductivity and
maximal pair condensation. The optimal doping point (corresponding
to the maximum of the SC pairing gap) has been used extensively in
the literature to mark the boundary between underdoped and
overdoped superconductors. It is the doping point where the
competition between the AF and SC correlations leads to the
maximal $T\tsub c$, but our results indicate that it does not mark
a boundary between qualitatively different physical regimes
characterized by qualitatively different wavefunctions.

According to the SU(4) model, the experimentally observed location
of the critical doping point can set a strict constraint on the
relative strengths of the three elementary interaction strengths,
$\chi$, $G_0$, and $G_1$. As one can see from Eq.\ (\ref{xq}),
$\chi$ must be either greater or less than the other two strengths
$G_0$ and $G_1$; otherwise $(\chi-G_0)/(\chi-G_1)$ would become
negative and no $x_q$ could occur in the physical range $0\le x\le
1$. Observations require that $\chi$ be greater than both $G_0$
and $G_1$ in order for the normal state at the half-filling to be
an AF state (Mott insulator).  It then follows that $G_0$ should
be intermediate in strength between $\chi$ and $G_1$; otherwise
$(\chi-G_0)/(\chi-G_1)>1$ will lead to $x_q>1$, which is outside
the physical doping range. Thus the SU(4) symmetry and the basic
experimental observations require that the three interaction
strengths must satisfy the condition $\chi > G_0 > G_1$. This is
the condition that we have assumed in (\ref{condition}).

It is instructive to examine two extreme cases of Eq.\ (\ref{xq}):
one is $x_q=1$ if $G_0 = G_1$; the other is $x_q=0$ if $\chi =
G_0$. For these two cases no critical doping point $x_q$ exists
within the physical doping range.  These cases correspond,
respectively, to the SO(4) and SO(5) symmetry limits of the SU(4)
model, as has been mentioned in Section II.\ A. For the former
case ($x_q=1$), there is no room for the SC phase at any $x$,
suggesting that the system in the entire doping range is in the AF
phase. For the latter case ($x_q=0$), the system is in the SO(5)
limit. Although the energy minimum in this limit is at
$\Deltaq=0$, the same as for the SC case, it is very shallow and
the system has large AF and pairing fluctuations. The SO(5) energy
surface at $x=0$ is in fact completely flat, develops an energy
minimum with $\Deltaq=0$ as the doping $x$ increases, and
gradually evolves to an SC-like state at $x=1$. This behavior is
characteristic of a critical dynamical symmetry, as discussed
extensively in Ref.\ \cite{lawu99}.

\subsubsection{Nature of the pseudogap}

The observation of pseudogap behavior in cuprates
\cite{Ti99,Shen03} has motivated a variety of theoretical
discussions but there is little agreement on the source of this
behavior. Two alternative classes of explanation for the origin of
pseudogaps have received considerable attention \cite{Shen03}:
(1)~The {\em preformed pair picture} \cite{EK95}, which suggests
that the formation of pairs and the condensation of those pairs
into a state with long-range order happen on two different energy
scales, with the pseudogap being the energy needed to form pairs
before pairing condensation.  If the pseudogap originates in
preformed pairs, it should merge with the pairing gap in the
overdoped region. (2)~The {\em competing-order picture}
\cite{Ta01}, which suggests that the pseudogap is an energy scale
associated with a form of order that competes with
superconductivity. If the pseudogap originates in competing order,
it should not merge with the pairing gap but should instead drop
to zero at a critical doping point where the competing order is
completely suppressed relative to the superconducting order.

In the solution of the SU(4) gap equations, the resulting energy
gaps (scales) are $\Delta_d$ (as well as $\Delta_0$),
$\Delta_\pi$, and $\Delta_q$. Generally the gaps $\Delta_d$ and
$\Delta_0$, may be interpreted as SC gaps and the gaps
$\Delta_\pi$ and $\Delta_q$ may be interpreted as pseudogaps.
However, our results suggest (as mentioned in Section IV.A and
discussed further below) that the pure spin-triplet state can
neither become a physical ground state at $T=0$, nor be reached by
thermal excitations.  The $\Delta_\pi$ gap can coexist with
$\Delta_q$ and $\Delta_d$ only in underdoped compounds and cannot
survive above $T\tsub c$.  Therefore, within the SU(4) model only
the AF gap $\Delta_q$ is a candidate for the observed pseudogap.

In the competing order picture, the energy scale that competes
with superconductivity in cuprates is often identified with
antiferromagnetism \cite{Ta01}, and this energy scale may be
called a pseudogap. The pseudogap energy scale vanishes at the
critical doping, which is independent of the disappearance of the
SC gap at $P\approx0.25$. Furthermore, experiments have shown (for
example, Ref.\ \cite{Ta01}) that the observed pseudogap is
intimately connected with the AF correlations, which disappear
exactly at the same critical doping point. The AF gap $\Deltaq$
has precisely these properties and thus may be interpreted as a
pseudogap. Specifically, the scale $\Deltaq$ in the SU(4) model is
the energy per electron required to break the AF correlation.
Therefore, the nature of the pseudogap in the SU(4) interpretation
is consistent with the competing order picture.

In the SU(4) model the spin-triplet pairing plays a mediating role
in the AF--SC competition. As can be seen from Fig.\ 2, its order
parameter $\Delta_\pi$ is zero at $x=0$ where the AF correlations
dominate, increases as doping increases to a maximum at half of
the critical doping value where the AF and SC correlations are
competing strongly, and finally disappears at the critical doping
point $x_q$ where the AF--SC competition ends.

On the other hand, the SU(4) model does not contradict the general
preformed pair picture conceptually. If we define preformed pairs
as pairs that are formed before pairing condensation, pairs with
only AF correlations may be interpreted as preformed pairs.
(These AF pairs correspond to a mixture of $D$ and $\pi$ pairs in
our Hilbert space.) This conclusion will become clear from the
discussions later in Section V. Thus $\Delta_q$ may be viewed also
as the scale of stabilization energy associated with preformed
pairs that condense into a pure SC pair state only after the AF
correlations are completely suppressed by increasing hole doping.
A pseudogap arising from this preformed pair picture can exist
only in the underdoped regime since $\Delta_q$ vanishes at the
critical doping point. For overdoped systems ($ x > x_q$) there is
no pseudogap and thus there are no preformed pairs in the gap
solutions found here.

\subsubsection{Scaling property of the gaps}

It is well known that pairing gaps in cuprates scale with $(T\tsub
c)\tsub{max}$ (the maximum superconducting transition temperature)
for all high-$T\tsub c$ compounds studied so far \cite{Ta95}.
According to Ref. \cite{Ta01}, there is also strong experimental
evidence indicating that the doping value of the putative quantum
critical point is universal in the hole-doped cuprates.

In Fig.\ \ref{fig:fg 1}, all the energy gaps are scaled by
$G_0\Omega$.  From Eq.\ (\ref{udopDd}), we know that for any given
$G_0$ the gap $\Deltad$ depends only on $x_q$ . Therefore, if the
critical doping point has a universal value the doping dependence
of $\Deltad$ should also be universal, since we shall show in the
next section that $(T\tsub c)\tsub{max}$ is proportional to $G_0$.
The gap $\Delta_0$ also has this scaling property because,
according to Eq.\ (\ref{odopDd}), $\Delta_0$ in the whole doping
range depends only on $G_0$.

However, the other two gaps $\Deltaq$ and $\Deltapi$ do not scale
in this way because they depend on $\chi/G_0$ and $G_1/G_0$,
respectively. Changing the ratio of $\chi/G_0$ and $G_1/G_0$ can
change the size of these gaps; thus the doping dependence of
$\Deltaq$ and $\Deltapi$ for different compounds generally could
be different. Only if the strength of triplet pairing $G_1$ is
zero (thus $\Deltapi=0$), can $\Deltaq$ be scaled because in this
case
$$
\chi/G_0=1/(1-x_q^2).
$$
Hence, the scaling property of $\Deltaq$ may be taken as an
indicator of triplet pairing strength in cuprates. Recent
$\Deltaq$ data \cite{Ta01,Kr02} seem to support such a scaling
property, at least approximately, implying that $G_1/G_0$ should
be small.

\section{Solution of gap equations at finite temperatures}

The gap equations for $T>0$ differ from those at $T=0$ in that the
terms $w_\pm$ (see Eq.\ (\ref{wpm})) acquire a temperature
dependence
\begin{equation}
w_\pm =\frac{P_\pm(T)}{e_\pm}=\frac{\tanh(R e_{\pm} /2k\tsub
BT)}{e_{\pm}}.
\label{wpmT}
\end{equation}
For finite temperature the gap equations could have a variety of
solutions, even for a fixed doping $x$. Which solution should
correspond to the physical ground state depends on temperature and
doping, and is determined by minimizing the energy. The solutions
that we have derived in Appendix C are obtained under the
(physically motivated) condition $\chi>G_0>G_1>0$. Under other
conditions the gap equations could have different solutions and
one has to solve Eqs.\ (\ref{gap:whole} -- \ref{1.22}) for
individual cases.

By comparing the energy densities expressed in Eq.\ (\ref{eqED}),
one can then find the solutions having the lowest energy and
determine the physical ground-state solution for given temperature
$T$ and doping $x$. We examine several cases below.

\subsection{The $\Deltad+\Deltaq+\Deltapi$ case}

This case corresponds to the all-gap solution found at $T=0$ for
the doping range $0\leq x\leq x_q$.  For $T>0$, the temperature
and doping dependent gap solutions are derived in Appendix C, with
the results
\begin{subequations}
\label{Tgap:whole}
\begin{eqnarray} \Deltaq&=&\frac
{\chi\Omega}{2} \sqrt{(x_q^{-1}-y)(x_q-y)}\ \frac xy
\label{eqDqyT}
\\
\Deltad&=& \frac{G_0\Omega}{2} \sqrt {x(x_q^{-1}-y)}\, g(y)
\label{eqDdyT}
\\
\Deltapi&=& \frac{G_1\Omega}2 \sqrt {x(x_q-y)} \, g(y)
\label{eqDpiyT}
\\
\lambdaPrime &=&-\frac{(\chi-G_1)\Omega}{2}
x_q\left(\frac{x}{y}-x_qx\right)-\frac{G_1\Omega}2x \label{eqLyT}
\end{eqnarray}
\end{subequations}
where $y$ and $g(y)$ are defined through
\begin{eqnarray}
y&=&\frac {x}{\sqrt{I_+(T)+I_-(T) \Gamma(y)}}\
\nonumber
\\
g(y)&=&\sqrt{\frac xy+\frac{I_+(T)-(x/y)^2}{2x(\bar{x}_q-y)}}
\nonumber
\end{eqnarray}
with
\begin{eqnarray}
&&\Gamma(y)=\frac{\bar{x}_q-y}{\sqrt{(x_q-y)(x_q^{-1}-y)}}
\nonumber\\
&&I_\pm(T)=\frac {P_-^2(T)\pm P_+^2(T)}2
\nonumber
\end{eqnarray}
and $\bar{x}_q\equiv (x_q^{-1}+x_q)/2$. It can be verified that
these solutions satisfy the SU(4) invariant
\begin{equation}
\langle{\cal E}\tsub{SU4}\rangle =\frac{\Omega^2}{4}
\left[I_+(T)-x^2\right].
\label{eqSU4T}
\end{equation}
With increasing temperature the system begins to break pairs,
with the number density of unpaired particles at temperature $T$
given by
\begin{equation}
 u =1-\sqrt{I_+(T)}=1-\sqrt{\frac{P^2_-(T)+P^2_+(T)}2}.
\label{equT}
\end{equation}

More specifically, to obtain the energy gaps for given doping $x$
and temperature $T$, one may adopt the following procedure. From
Eqs.\ (\ref{wpmT}) and (\ref{qpe}) one obtains
\begin{equation}
T=\frac{R \sqrt{ (\Deltaq\pm\lambdaPrime)^2+{\Delta_\pm}^2}}
{2k\tsub B\,\atanh(w_\pm e_\pm)}\ .
\label{3.6}
\end{equation}
This equation implies that
\begin{equation}
\frac{\sqrt{(\Deltaq+\lambdaPrime)^2+{\Delta_+}^2}}{\atanh(w_+e_+)}
=\frac{\sqrt{(\Deltaq-\lambdaPrime)^2+{\Delta_-}^2}}{\atanh(w_-e_-)}.
\label{3.5}
\end{equation}
By solving Eqs.\ (\ref{subgapeq:q} -- \ref{subgapeq:la}) directly,
one gets
\begin{equation}
w_\pm= \frac{\frac{2\Deltaq}{\chi\Omega}\mp x}
{\Deltaq\pm\lambdaPrime}.
\nonumber
\end{equation}
Now by using Eq.\ (\ref{A2.6}) in Appendix C to convert $\Deltapi$
into $\Deltad$
\begin{equation}
 \Deltapi =\left (\frac{w_-
-\omega_0}{w_--\omega_1}\right )\Deltad,
\label{DpitoDd}
\end{equation}
$\Delta_+$ and $\Delta_-$ can be related to $\Deltad$
\begin{equation}
 \Delta_\pm =\left [1 \pm\left (\frac{w_-
-\omega_0}{w_--\omega_1}\right )\right ]\Deltad.
\label{DpmtoDd}
\end{equation}
Thus, for a given doping $x$, one can solve for $y$ from Eq.\
(\ref{eqDqyT}), and $\lambdaPrime$ from Eq.\ (\ref{eqLyT}), for
each $\Deltaq$.  The pairing gap $\Deltad$ can then be obtained by
solving Eq.\ (\ref{3.5}) directly. With $\Deltad$ determined,
$\Deltapi$ and the corresponding temperature $T$ can be calculated
from Eqs.\ (\ref{DpitoDd}) and (\ref{3.6}), respectively.

\subsection{The $\Deltad$ case}

This case corresponds to one of the trivial solutions at $T=0$.
Following a similar procedure as in the $T=0$ case, one obtains
the temperature and doping dependent gap solutions
\begin{subequations}
\label{TgapDd:whole}
\begin{eqnarray} \Deltaq &=& \Deltapi = 0
\label{eqDpiq}
\\
\Deltad &=& \frac{G_0\Omega}{2} \sqrt{I_+(T)-x^2}
\label{eqDdG}
\\
\lambdaPrime &=&- \frac{G_0\Omega}{2}x.
\label{eqLT}
\end{eqnarray}
\end{subequations}
Note that in the present case,
$$
I_+(T)=P^2_+(T)=P^2_-(T).
$$
By
using Eqs.\ (\ref{qpe}), (\ref{wpmT}), and (\ref{eqLT}), Eq.\
(\ref{eqDdG}) may be written in the following form
\begin{equation}
T=\frac{R \sqrt{\left(\dfrac{G_0\Omega}2x\right)^2+{\Deltad}^2} }
{2k\tsub B\ \atanh\left[
\sqrt{\left(\dfrac{2\Deltad}{G_0\Omega}\right)^2+x^2}\,\right]}\ .
\label{3.8}
\end{equation}
Therefore, for given $x$ and $T$ the gap $\Delta_d$ can be
obtained from Eq.\ (\ref{3.8}), and $\lambdaPrime$ obtained from
(\ref{eqLT}).

Since now all the correlations are zero except the spin-singlet
pairing, we have
\begin{equation}
\langle{\cal E}\tsub{SU4}\rangle = \langle D^\dagger D\rangle =
\left(\frac{\Deltad}{G_0}\right)^2 .
\nonumber
\end{equation}
One can see that the SU(4) invariant (\ref{eqSU4T}) is also valid
in this case, and the number density of unpaired particles can be
evaluated using Eq.\ (\ref{equT}).  It can be checked without
difficulty that when $T\rightarrow 0$,
$$
P_\pm(T)\rightarrow 1 \qquad I_+(T)\rightarrow
1,
$$
and the solutions (\ref{TgapDd:whole}) then reduce to Eq.\
(\ref{gapDd:whole}) of the $T=0$ solution.

\subsection{ The $\Deltaq$ case}

In this case both $\Deltad$ and $\Deltapi$ are zero, corresponding
to a solution that applies only for temperatures $T>T\tsub c$:
\begin{subequations}
\label{TgapDq:whole}
\begin{eqnarray}
\Deltad &=& \Deltapi = 0 \\
\Deltaq &=&\frac{\chi\Omega}2(P_-(T)-x).
\label{eqDqPx}
\end{eqnarray}
\end{subequations}
These results can be derived from Eqs.\ (\ref{subgapeq:q}) and
(\ref{subgapeq:la}), which in the present case reduce to
\begin{eqnarray}
\frac{4\Deltaq}{\chi\Omega}&=&P_-(T)+P_+(T), \label{eqDqP}
\\
2x\ &=&P_-(T)-P_+(T).
\label{eqxP}
\end{eqnarray}
Using Eq.\ (\ref{eqxP}), Eq.\ (\ref{eqDqPx}) can be rewritten as
\begin{equation}
\Deltaq =\frac{\chi\Omega}2 \sqrt{I_+(T)-x^2}.
\nonumber
\end{equation}
In a manner similar to that for the $\Deltad$ case, one can show
that Eqs.\ (\ref{eqSU4T}) and (\ref{equT}) are also valid for the
pure AF case ($\Deltad=\Deltapi=0$).  Thus Eqs.\ (\ref{eqSU4T})
and (\ref{equT}) are actually general expressions; for different
cases only the values of the quasiparticle energies $e_\pm$
(contained in $P_\pm(T)$) differ.

By solving Eqs.\ (\ref{eqDqP}) and (\ref{eqxP}) one obtains
\begin{eqnarray}
T&=&\frac{R \Deltaq }{k\tsub B A_+}
\label{3.9}
\\
\lambdaPrime &=&\frac{k\tsub B T A_-}{R},
\label{3.10}
\end{eqnarray}
with
$$
A_\pm \equiv \left[\atanh\left(\dfrac{2\Deltaq}{\chi\Omega}-x\right)
\pm \atanh\left(\dfrac{2\Deltaq}{\chi\Omega}+x\right)\right ] .
$$
Thus, for a temperature $T$, $\Deltaq$ can be obtained from Eq.\
(\ref{3.9}) and $\lambdaPrime$ from Eq.\ (\ref{3.10}).

There is an intriguing point concerning the pure AF state at $T=0$
that should be clarified.  On one hand, when $T=0$, we have that
$P_-(0) =1$ and $P_+(0) =1-2x$. Thus
$$
I_+(0) = 1-2x+2x^2,
$$
and according
to Eq.\ (\ref{equT}), $u>0$. This means that, if $x\ne 0$, the pure
AF phase requires pair breaking even at $T=0$. The SU(4) invariant
in this case is
$$
\langle{\cal E}\tsub{SU4}\rangle = \langle
Q^\dagger Q\rangle = \frac{\Omega^2}{4} \left(1-x\right)^2.
$$
On the other hand, it is our basic assumption that when $T=0$ all
particles are paired $(u=0)$ and the SU(4) invariant should be
$$
\langle{\cal E}\tsub{SU4}\rangle = \frac{\Omega^2}{4}
\left(1-x^2\right),
$$
which is state-independent if SU(4) symmetry holds.  Despite
appearances, these two expressions for $\langle{\cal
E}\tsub{SU4}\rangle$ are not in contradiction because when there
exist pairing correlations (one or both of $G_0$ and $G_1$
nonzero), no AF state can be the $T=0$ ground state unless $x=0$.
The only possibility for the pure AF state to become the ground
state at $T=0$ for $x>0$ is in the absence of pairing
($G_0=G_1=0$). But under this condition it is not necessary to
require $\langle D^\dagger D\rangle = \langle \pi^\dagger
\pi\rangle=0$, so that $\langle D^\dagger D\rangle$ and $\langle
\pi^\dagger \pi\rangle$ can supplement $\langle Q^\dagger
Q\rangle$ to preserve the invariance $\langle{\cal
E}\tsub{SU4}\rangle = \frac{\Omega^2}{4} \left(1-x^2\right)$.

\section{The phase diagram}

We now use the preceding results to construct temperature versus
doping phase diagrams in the SU(4) model. The interaction strength
parameters of the theory represent effective interactions within a
truncated space.  For an effective interaction approach to make
sense, the interaction strengths cannot have very strong local
dependence on global parameters such as doping fraction (except
for possible rapid changes at phase boundaries). However, since
the effective interaction strengths represent renormalized
interactions within a highly truncated space, they may reasonably
be expected to have a smooth dependence on the global parameters.
Nevertheless, it is useful to make the simplest assumption for a
starting point: the effective interaction strengths are {\em
constant} across the entire doping range relevant to cuprates.
Although this is a more stringent restriction than is warranted
for a realistic theory, it has the advantage of simplicity and it
should give at least sensible qualitative results if the present
approach is valid.

\epsfigbox{fg2}{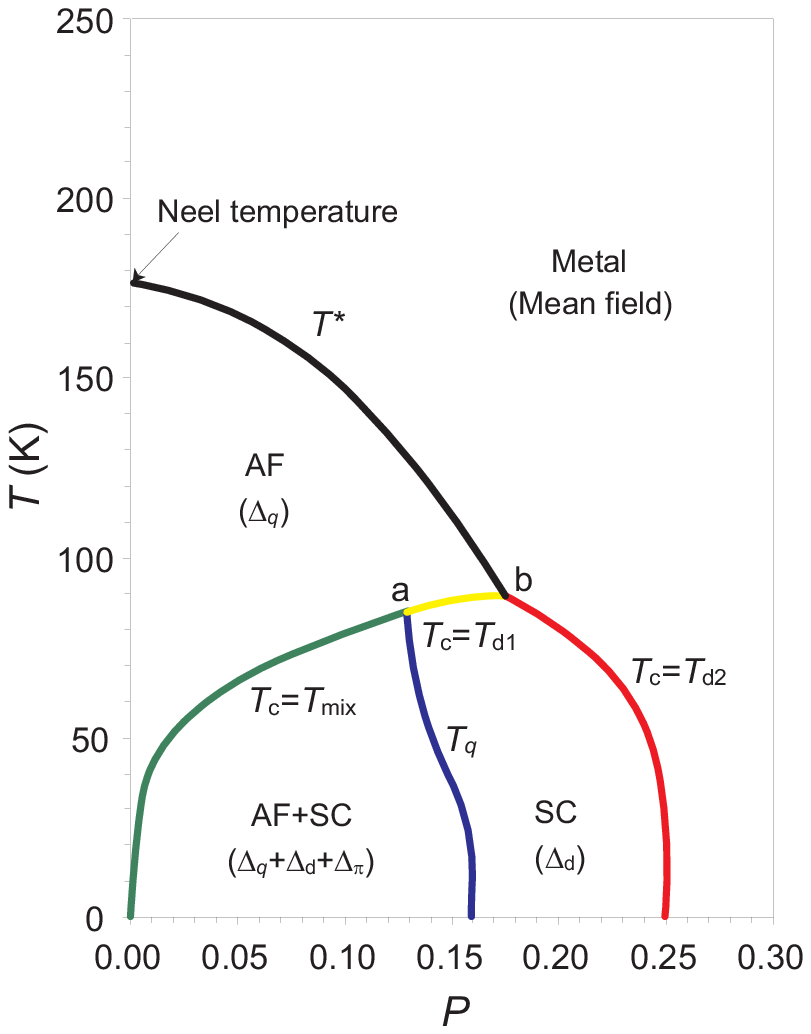}{0pt}{0pt} {(Color online) Phase
diagram predicted by the SU(4) model with $R=0.6$. The interaction
strengths are the same as those used in Fig.\ \ref{fig:fg 1}, but
in units of $k_B(T\tsub c)\tsub{max}$, where $(T\tsub
c)\tsub{max}$ is taken to be 90 K.
The two points marked as $a$ and $b$ are tricritical points. The
critical doping point is at $P_q=0.16$ (corresponding to
$x_q=0.64$).}

In Fig.\ \ref{fig:fg2} we show a typical phase diagram. In the
calculations the interaction strengths are kept constant, with the
same values as those in the zero-temperature case discussed in
Figs.\ 1 and 2. The only adjustable parameter is $R$ (appearing in
Eq.\ (\ref{thermal})), the energy scaling parameter that
approximately corrects for the possible error caused by the
degeneracy assumption in the quasiparticle excitation spectrum for
a finite $T$.

There are four distinct phases emerging in Fig.\ \ref{fig:fg2}: a
pure antiferromagnetic phase (AF), a superconducting phase (SC), a
transitional phase with all three correlations present, which may
be called a mixed phase (marked as AF+SC), and a metallic phase.
The correlations (energy gaps) associated with each phase are
indicated in parentheses. The doping-dependent transition
temperatures $T\tsub c$, $T^*$, and $T_q$ define the boundaries
for these phases. There are two tricritical points $a$ and $b$,
which are the intersection points of three phases.

\subsection{Phases below the critical temperature}

We first discuss the phases below the critical temperature $T\tsub
c$. In the phase diagram of Fig.\ 3 there are two phases below
$T\tsub c$, the mixed phase and the SC phase. The essential
difference between them is that the mixed phase has both AF
correlation $\Deltaq$ and triplet pairing $\Deltapi$ mixed with
the singlet pairing $\Deltad$, while the SC phase has a pure
$\Deltad$ correlation with $\Deltaq=\Deltapi= 0$. At zero
temperature, the separation point between these two phases
corresponds to the critical doping point $x_q$.

The boundary between the mixed phase and the SC phase is marked in
Fig.\ 3 as $T_q$, which is the locus of all points where the
corresponding energies of the two phases are equal
\begin{equation}
\frac{[\Deltad^{(2)}(T_q)]^2}{G_0}= \frac{[\Deltad^{(1)}(T_q
)]^2}{G_0}+\frac{[\Deltapi^{(1)}(T_q )]^2}{G_1}
+\frac{[\Deltaq^{(1)}(T_q)]^2}{\chi}.
\label{5.1}
\end{equation}
Here we use the superscript indices `(1)' and `(2)' to distinguish
quantities evaluated in the mixed and SC phases, respectively.

Eq.\ (\ref{5.1}) indicates that the order parameters $\Deltad$,
$\Deltapi$, and $\Deltaq$ could be discontinuous across the
boundary except for $T=0$. This implies that the mixed--SC phase
transition is second-order at $T=0$, but could generally be
first-order for  finite $T$.  The actual situation depends on the
relative location of the critical points $a$, $b$, and $x_q$.
Using $P_\sigma$ (or $x_\sigma$) with $\sigma=a$ or $b$ to
represent doping fractions corresponding to the tricritical
points, we can show that the transition is indeed first-order if
$x_q< x_b$, as illustrated in Fig.\ 3.

Quite interestingly, if $x_q\ge x_b$ the point $a$ merges with the
point $b$, as illustrated in Fig.\ \ref{fig:fg3}. The point $b$ is
now a quadcritical point, where at this unique doping and
temperature {\it all four} phases that we have identified in the
SU(4) model can coexist in equilibrium with each other. In this
case, all the order parameters change smoothly across the boundary
and the mixed--SC phase transition becomes second-order. The curve
$T_q$ in this case looks like an extension of the curve for
transition temperature $T^*$. The results in Fig.\ 4 are obtained
for a larger $x_q$ value than for those in Fig.\ 3. This suggests
that the precise value of the critical doping point is related to
the number of tricritical points in the phase diagram and to the
order of the phase transitions.

Both results presented in Figs.\ 3 and 4 are possible solutions.
The only difference between them is the triplet pairing strength
$G_1$. In Fig.\ 3, $G_1=1.3$ is used, resulting in a smaller
$x_q=0.64$, while in Fig.\ 4, $G_1=4.7$ corresponding to
$x_q=0.76$. The former case has $x_q<x_b$ and the latter
$x_q>x_b$. This suggests that the relative position of $x_q$ and
$x_b$ could be an indicator of the strength of triplet pairing. It
would be extremely interesting to confirm experimentally whether
the real phase diagram in cuprates is of the Fig.\ 3 or Fig.\ 4
type.

\epsfigbox{fg3}{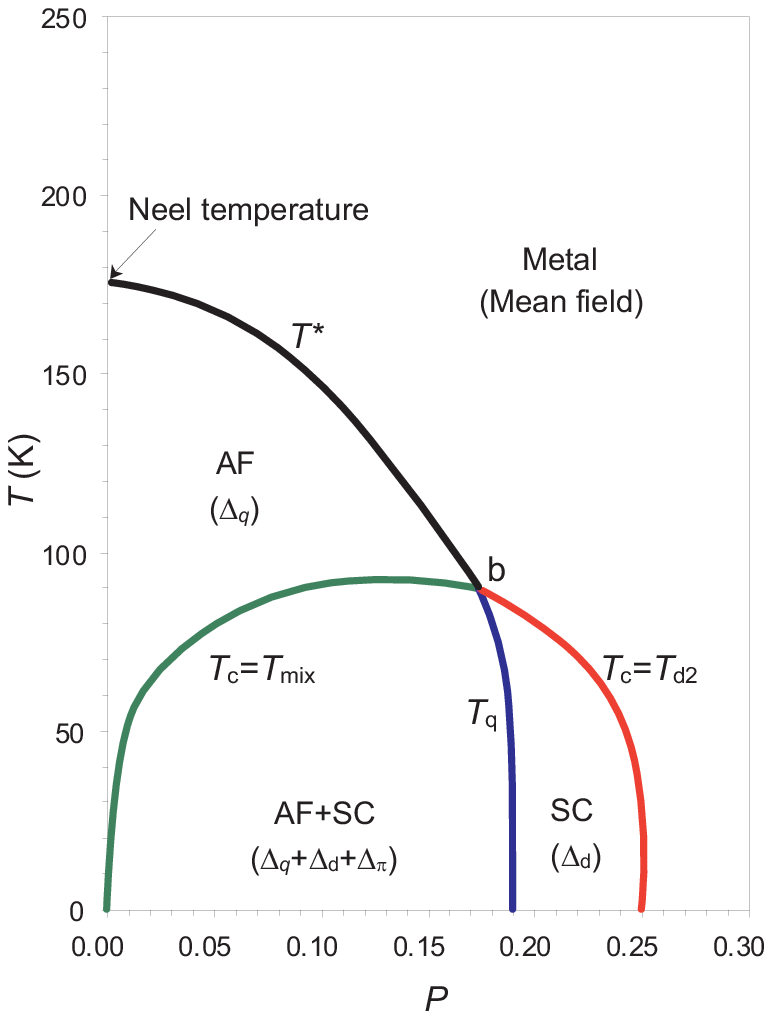}{0pt}{0pt} {(Color online) Phase
diagram predicted by the SU(4) model for $P_q>P_b$. $P_q=0.19$
(corresponding to $x_q=0.76$) is chosen in this figure while
$P_b=0.175$ (corresponding to $x_q=0.7$) is the same as that in
Fig.\ \ref{fig:fg2}. All the parameters remain unchanged except
that $G_1$ increases from 1.3 to 4.7 (in units of $k_B({T\tsub
c})\tsub{max}$) for a larger $P_q$ value. }

\subsection{The critical temperature}

Let us use Fig.\ \ref{fig:fg2} to discuss the critical temperature
$T\tsub c$. The curve for $T\tsub c$ in the SU(4) phase diagram
consists of three segments:
$$
T\tsub c = \left \{ \begin{array}{ll}
T\tsub{mix} &\quad (P\le P_a)
\\
T\tsub{d1} &\quad (P_a< P < P_b)
\\
T\tsub{d2} &\quad (P\ge P_b)
\end{array}
\right. ,
$$
which are seen to behave quite differently. The discussions below
apply also to Fig.\ \ref{fig:fg3} because in this case there is
only one phase boundary intersection
 point $b$ and thus the segment $T\tsub{d1}$ does not
exist.

In the doping range $P\ge P_b$ the critical temperature is $T\tsub
c=T\tsub{d2}$, which is the SC--metal transition temperature and
can be expressed analytically using  Eq.\ (\ref{3.8}) with
$\Deltad=0$:
\begin{eqnarray}
T\tsub c &=& T\tsub{d2}=T({\rm SC}\leftrightarrow {\rm metal})
\nonumber
\\
 &=& G_0\Omega\frac{Rx}{4k\tsub B \, \atanh \left(x\right)}
\qquad (x \ge x_b).
\label{5.2}
\end{eqnarray}

In the doping range $P\le P_a$ the critical temperature is
$T\tsub c=T\tsub{mix}$, which is the mixed--AF transition
temperature determined from Eq.\ (\ref{3.6}) with $\Deltad=0$
(thus $\Delta_\pm=0$):
\begin{eqnarray}
T\tsub c &=& T\tsub{mix}=T({\rm mixed} \leftrightarrow {\rm AF})
\nonumber
\\
&=& \frac{R\,|\Deltaq\pm\lambdaPrime|} {2k\tsub B
\, \atanh \left(\dfrac{2\Deltaq}{\chi\Omega}\mp x\right )}
\qquad (x\leq x_a) ,
\label{5.3}
\end{eqnarray}
where $\lambdaPrime$ depends on $\Deltaq$ through $y$ [\,see Eqs.\
(\ref{eqDqyT}) and (\ref{eqLyT})\,] and $\Deltaq$ is determined by
Eq.\ (\ref{3.5}).

In the doping range $P_a<P<P_b$ the critical temperature is
$T\tsub c=T\tsub{d1}$, which is the SC--AF phase boundary and
determined by the condition that at $T\tsub c$ the energies of the
two phases are equal:
\begin{equation}
\frac{[{\Deltad(T\tsub c)}]^2}{G_0} =\frac{[{\Deltaq}(T\tsub
c)]^2}{\chi}.
\label{5.4}
\end{equation}
Using this condition and Eqs.\ (\ref{3.8}) and (\ref{3.9}), one
can obtain the pairing gap $\Deltad$ through the following
equation
\begin{equation}
\begin{array}{l}
\displaystyle \frac{
              \sqrt{
                   \left(\dfrac{G_0\Omega}{2}x
              \right)^2+{\Deltad}^2
              }
              }{2 \, \atanh \left[
                               \dfrac{2}{G_0\Omega}
              \sqrt{
                   \left(\dfrac{G_0\Omega}{2}x
                   \right)^2+{\Deltad}^2
              }
                          \right]
              } =
\\[4em]
\displaystyle
\frac{
         \sqrt{
              \chi/G_0
         }
         \Deltad
     }
     {
         \atanh \left(\dfrac{2\Deltad}{
                \sqrt{
                      \chi G_0
                }
                \Omega}-x\right)
                +\atanh \left(\dfrac{2\Deltad}{
                \sqrt{
                      \chi G_0
                }
                \Omega}+x\right)
     }
\end{array}
\nonumber
\end{equation}
and $T\tsub c$ is then determined by the equation
\begin{eqnarray}
T\tsub c &=& T\tsub{d1}= T({\rm SC}\leftrightarrow {\rm AF})
\nonumber
\\
&=& \frac{R \sqrt{
                   \left(\dfrac{G_0\Omega}{2}x
                   \right)^2+{\Deltad}^2
            }
          }
          {2 k\tsub B\
             \atanh \left[\sqrt{
                  \left(\dfrac{2\Deltad}{G_0\Omega}
                                     \right)^2+x^2}
                          \right]
              }  ,
\label{eqTd1}
\end{eqnarray}
which is valid in the doping range $x_a\leq x \leq x_b$.

The phase transitions in the two regions ($P\le P_a$ and $P\ge
P_b$) are second-order (all gaps change smoothly across the
boundary). However, transitions in the range $x_a\leq x \leq x_b$
behave very differently. The pairing gap $\Deltad$ in this doping
range drops discontinuously to zero when crossing the boundary,
while the AF gap $\Deltaq$ jumps from zero to a finite value,
suggesting that the system undergoes a first-order phase
transition from the SC to AF phase.

In traditional BCS theory the energy gap appearing in the density
of states is identified with the pairing gap.  This remains true
for the overdoped regime ($T\tsub{d2}$) where the singlet pairing
is the only correlation. However, this is {\em not true} for
underdoped systems where more correlations are involved and a
pairing gap is not necessarily equivalent to the energy gap
appearing in the density of states. As shown in Eq.\ (\ref{qpe}),
in the underdoped regime with the mixed phase there are two gaps,
$\Deltapm$, in the density of states, but neither of them is an
order parameter. The actual process of the mixed--AF transition is
as follows: with increasing temperature, $\Deltapi\rightarrow 0$
at a temperature that causes $w_-=w_0$ before $T\tsub c$ is
reached [see Eqs.\ (\ref{DpitoDd} -- \ref{DpmtoDd})], and at this
temperature $\Deltapm$ reduce to $\Deltad$. After that, the order
parameter $\Deltad\rightarrow 0$ at $T\tsub c$.

This complexity in gap structure suggests that one has to be
careful with the interpretation of energy gaps measured in
underdoped compounds. According to the SU(4) model, there are
several gaps involved in this doping and temperature range such as
$\Deltad$, $\Delta_0$, $\Deltapi$, $\Delta_+$, and $\Delta_-$, and
they all can correspond to energy scales having the same order of
magnitude. Therefore, an energy gap by two different experimental
probes may not really be the same gap since a particular
measurement may be sensitive only to certain of these gaps because
of their microscopic structure.

The location of the tricritical point $x_a$ can be determined
from $T\tsub{mix}(x_a)=T\tsub{d1}(x_a)$ with the condition that
the energies of the mixed and SC phases at this point are
equivalent. By using Eqs.\ (\ref{5.3}) and (\ref{eqTd1}), $x_a$
can be obtained by solving the following equations
\begin{equation}
\frac{|\Deltaq\pm\lambdaPrime|} { \, \atanh
\left(\dfrac{2\Deltaq}{\chi\Omega}\mp x_a\right )}=\frac{\sqrt{
                   \left(\dfrac{G_0\Omega}{2}x_a
                   \right)^2+{\Deltad}^2
            }
          }
          {\atanh \left(\sqrt{ \left(\dfrac{2\Deltad}{G_0\Omega}
                                     \right)^2+x_a^2}
                          \right)
              }
\nonumber
\end{equation}
where $\Deltaq$ and $\lambdaPrime$ are known if $T\tsub{mix}$ is
determined, while $\Deltad$ is calculated through the condition
$\Deltaq^2/\chi=\Deltad^2/G_0$.

\subsection{Phases above the critical temperature}

For temperatures above $T\tsub c$ there are two possible phases:
the pure AF phase and the metallic phase.

As already mentioned, the pure AF phase at $T=0$ can exist only at
$x=0$ as a limit of the mixed phase. This is because of pairing
correlations that give the mixed phase a more favorable energy,
and is qualitatively consistent with the observation that the AF
phase is favored only in a very restricted doping region near half
filling.

When $T>T\tsub c$, however, the pure AF phase can become the
ground state in the underdoped regime, as predicted in Fig.\
\ref{fig:fg2}. This is because thermal fluctuations can destroy
the pairing correlations preferentially relative to the AF
correlations. If temperature increases further, the AF correlation
will eventually be totally destroyed ($\Deltaq\rightarrow 0$) and
the system will undergo an AF--metal transition.

The AF--metal transition is second-order.  The transition
temperature $T^*$ is determined by Eq.\ (\ref{3.9}) in the limit
that $\Deltaq\rightarrow 0$:
\begin{eqnarray}
T^* &=& T({\rm AF}\leftrightarrow {\rm metal})
\nonumber
\\
&=& \frac{R\chi\Omega(1-x^2)}{4k\tsub B} \qquad (0\le x\le x_b).
\label{5.17}
\end{eqnarray}
The N\'eel temperature can be determined from the preceding
equation with $x= 0$,
\begin{equation}
T\tsub N=
\frac{R\chi\Omega}{4k\tsub B } \qquad (x=0).
\nonumber
\end{equation}

The location of tricritical point $x_b$ can be determined through
the condition $T\tsub{d2}(x_b)=T^*(x_b)$. Combining Eq.\
(\ref{5.17}) with (\ref{5.2}) gives
\begin{equation}
x_b = \sqrt{1-\frac{G_0}{\chi}\frac{x_b}{\atanh (x_b)} }.
\nonumber
\end{equation}

\subsection{Discrepancies}

The phase diagrams of the SU(4) model are consistent with much of
the current understanding for cuprate systems. However, there are
two obvious quantitative discrepancies between the simplified
results presented here and data.

One discrepancy occurs at very low doping. Data show that the pure
AF phase is not confined to half filling but can extend over a
narrow non-zero doping range at low temperature. Experimentally,
$T\tsub{c}$ goes to zero at $P\approx 0.05$, which differs from
the theoretical prediction $P=0$. However, recall that we have
deliberately employed an oversimplified model here (effective
interactions independent of doping) in order to emphasize that the
qualitative features of the SU(4) quasiparticle solution follow
from the physics encoded in the algebraic structure, not from
detailed parameter adjustment.  As we have shown, in the SU(4)
model the condition for the pure AF state to be the ground state
is the complete absence of pairing interactions.  Since we have
assumed all effective interaction strengths to be constant over
the whole doping range, the pairing correlation exists at all
dopings.  This is not a favorable condition for the AF phase to be
the ground state and with these assumptions $T\tsub c$ can go to
zero only at the both ends of the doping range, $x=0$ ($P=0$) and
$x=1$ ($P=0.25$). This discrepancy suggests that there exists an
onset of effective pairing correlation around $P_i\approx 0.05$.
If we allow a simple variation of pairing strength with doping
implying that the pairs become stable and thus have pairing
correlation only when $P>P_i$, this discrepancy can be resolved
easily.

A second quantitative discrepancy is that our predicted N\'eel
temperature is too low (175 K) for a cuprate system with
$(T\tsub{c})\tsub{max}\approx 90$ K.  Again, if we relax the
assumption of constant interaction strength by considering
different AF correlation strength $\chi$ before and after the
onset of pairing near $P=0.05$ (for example, maximal $\chi$ at
$x=0$ that decreases as doping increases and finally is stabilized
at a smaller value when pairing is established), this problem is
also easy to resolve.  Work to establish the mechanism for pair
formation and thereby to quantify the expected doping dependence
of the effective interactions is in progress and details will be
reported elsewhere.

\section{Summary}

The present work has used dynamical symmetries, Lie algebras, and
generalized coherent states to derive the temperature-dependent
gap equations expected for a theory in which antiferromagnetism
and $d$-wave superconductivity compete on an equal footing, and
for which the normal undoped states have Mott insulator character.
Although the SU(4) model is constructed using symmetry principles,
the coherent state method permits the problem to be cast in the
form of a generalized quasiparticle problem. Therefore, the
results are presented in terms of equations that may be recognized
as a generalization of the BCS formalism to include more than one
kind of pairing, subject to an SU(4) symmetry constraint. This
symmetry constraint has a clear physical origin.  As the preceding
discussion has shown, SU(4) symmetry may be viewed as concise
shorthand for competing antiferromagnetism and $d$-wave
superconductivity on a 2-dimensional lattice with no double
occupancy.

The quasiparticle structure that results for the cuprate
superconductors is rather rich, giving rise to novel gap
structures including both pairing gaps and pseudogaps, and a
potentially complex phase diagram. A critical doping point $P_q$
appears naturally in the theory as the boundary between doping
regimes having qualitatively different ground-state wavefunctions
at zero temperature: a pure superconducting solution at higher
doping ($P>P_q$) and a solution with superconductivity strongly
suppressed by antiferromagnetism at lower doping ($P<P_q$). Thus,
the critical doping point is associated with a quantum phase
transition.  The pairing gap has been shown to have two solutions
for $P<P_q$: a small gap, associated with competition between
superconductivity and antiferromagnetism that is responsible for
the ground-state superconductivity in underdoped systems, and a
large gap without antiferromagnetic suppression that corresponds
to a collective excited pairing state.

Within the SU(4) model a pseudogap can occur naturally. It has
been demonstrated that the pseudogap originates from the
competition of antiferromagnetism with $d$-wave superconductivity
in the underdoped regime, and terminates exactly at the critical
doping point. These conclusions are in accord with many current
observations in cuprates. Although the pseudogap arises directly
from competing antiferromagnetic and superconducting order, we
have also argued that it may be interpreted in a preformed pair
picture since the corresponding wavefunction contains singlet and
triplet pairs fluctuating in an antiferromagnetic background but
no long-range pairing order.

Once the parameters in the SU(4) gap diagram are determined by
fitting to the gap data, a rich phase diagram as a function of
temperature and doping can be sketched with only one additional
adjustable parameter. A variety of phases reflecting the interplay
of the spin-singlet and spin-triplet pairing with the
antiferromagnetic correlations has been predicted. Properties of
the phases may be expressed quantitatively using the mathematical
properties of the SU(4) algebra and its coherent states, leading
to a set of testable predictions concerning phase structure and
phase transitions in the cuprate superconductors.

Finally, we note that the striking properties in the phase diagram
discussed in this paper have similarities with properties observed
in other materials (often at lower temperature and energy scales).
For example, in heavy-fermion compounds and some 2D organic
superconductors a superconducting phase appears near the boundary
of an AF phase. As a second example, the manganites have complex
competing phases, some bearing a resemblance to those discussed in
this paper. Therefore, we expect that the general formalism
presented here, which is uniquely suited to deal quantitatively
with multiple competing low-temperatures phases in strongly
correlated systems, will prove applicable to a much broader range
of systems with doping replaced or supplemented by additional
control parameters such as pressure or strength of magnetic field.

\section*{Acknowledgments}

We are grateful for useful discussions with G. Arnold, E. Dagatto,
P.-C. Dai, T. Egami, B. Fine, T. K. Lee, A. Moreo, C.-Y. Mou, T.
Papenbrock, and H.-H. Wen.

\appendix
\section{Matrix Representation of the SU(4) Algebra}

To facilitate the evaluation of  $\delta\langle H'\rangle=0$ in
section III.A, it is convenient to use a matrix representation
introduced in Ref.\ \cite{lawu99,wmzha88}. In this representation
the SU(4) generators are expressed in terms of 4$\times$4
matrices, with the matrix elements defined in the following
4-dimensional single-particle basis:
\begin{equation}
\{ c^\dagger_{{\bf r}\uparrow}|0^*\rangle,\ c^\dagger_{{\bf
r}\downarrow}|0^*\rangle,\ c_{\bar{\bf r}\uparrow}|0^*\rangle,\
c_{\bar{\bf r}\downarrow}|0^*\rangle \}.
\nonumber
\end{equation}

Explicitly, for an operator $\hat O$ this is implemented through
the following mapping:
\begin{eqnarray}
\left[
\begin{array}{cc} O^{11} &  O^{12} \\  O^{21} &  O^{22}
\end{array}
\right]&\Rightarrow& \hat O= \sum_{{\bf r},i,j} \left[
O^{(11)}_{ij}c^\dagger_{{\bf r} i} c_{{\bf r}j}
+O^{(22)}_{ij}c_{\bar{\bf r} i} c^\dagger_{\bar{\bf r} j}
\right.\nonumber\\
&+& \left. O^{(12)}_{ij} c^\dagger_{{\bf r}i} c^\dagger_{\bar{\bf
r}j}
+O^{(21)}_{ij} c_{\bar{\bf r}i} c_{{\bf r} j } \ \right],
\label{mapping}
\end{eqnarray}
where $O^{kl}$ ($k,l=1$ or 2) are 2$\times$2 matrices with matrix
elements $O^{(kl)}_{ij}$. A 4-dimensional faithful matrix
representation of SU(4) generators (\ref{rspace2}) can be obtained
immediately from this mapping:
\begin{equation}
\begin{array}{llll}
{p}_{12}^{\dagger } = &\left[
\begin{array}{cc}0 &\ i\sigma _y \\ 0 &\
0\end{array}
\right] &
\quad
{p}_{12}^{} = & \left[ \begin{array}{cc} 0 &\
0 \\ -i\sigma _y &\ 0
\end{array}\right]
\\ [1em]
{q}_{12}^{\dagger } = &\left[
\begin{array}{cc} 0 &\ \sigma _x \\ 0 &\
0\end{array}
\right] &
\quad
{q}_{12}^{} = &\left[
\begin{array}{cc} 0 &\
0 \\
\sigma _x &\ 0
\end{array}
\right]
\\[1em]
{q}_{11}^{\dagger } = & \left[
\begin{array}{cc}0 &\ I+\sigma _z \\ 0 &\ 0
\end{array}
\right] &
\quad
{q}_{11}^{} = &\left[ \begin{array}{cc} 0 &\
0 \\ I+\sigma _z &\ 0
\end{array} \right]
\\[1em]
{q}_{22}^{\dagger } = &\left[
\begin{array}{cc}
0 &\  I-\sigma _z \\ 0 &\ 0
\end{array}
\right] &
\quad
{q}_{22}^{} = & \left[
\begin{array}{cc} 0 &\ 0 \\ I-\sigma _z &\ 0
 \end{array}\right]
\\ [1em]
{S}_{12}^{} = & \left[
\begin{array}{cc}
\sigma _{+} &\  0 \\ 0 &\  -\sigma _{-}
\end{array}
\right] &
\quad
{S}_{21}^{} = & \left[ \begin{array}{cc}
\sigma _{-} &\ 0 \\ 0 &\ -\sigma _{+}
\end{array}
\right]
\\[1em]
{S}_{11}^{} = &\left[
\begin{array}{cc}
\frac{I+\sigma _z}2 &\ 0 \\ 0 &\ -\frac{I+\sigma _z}2
\end{array}
\right] &
\quad
{S}_{22}^{} = & \left[ \begin{array}{cc}
\frac{I-\sigma _z}2 &\ 0 \\ 0 &\ -\frac{I-\sigma _z}2
\end{array}
\right]
\\ [1em]
\tilde{Q}_{12}^{} = & \left[
\begin{array}{cc}
\sigma _{+} &\ 0 \\ 0 &\ \sigma _{-}
\end{array}\
\right] &
\quad
\tilde{Q}_{21}^{} = &\left[ \begin{array}{cc}
\sigma _{-} &\ 0 \\ 0 &\ \sigma _{+}
\end{array}
\right]
\\[1em]
\tilde{Q}_{11}^{} = & \left[ \begin{array}{cc}
\frac{I+\sigma _z}2 &\ 0 \\ 0 &\ \frac{I+\sigma _z}2
\end{array}
\right] &
\quad
\tilde{Q}_{22}^{} = & \left[ \begin{array}{cc}
\frac{I-\sigma _z}2 &\ 0 \\ 0 &\ \frac{I-\sigma _z}2
\end{array}
\right]
\end{array}
\label{matrix}
 \end{equation}
where $\sigma _x$, $\sigma _y$, and $\sigma _z$ are Pauli matrices
in the standard representation, $\sigma _{\pm } \equiv \tfrac
12(\sigma _x\pm i\sigma _y)$, and $I$ is a unit matrix. A
corresponding matrix representation of the operators
(\ref{operatorset}) is then constructed readily from Eq.\
(\ref{matrix}). The unitary transformation operator ${\cal T}$ of
Eq.\ (\ref{eq10}) may be written in this matrix representation as
\begin{equation}
{\cal T} = \left[
\begin{array}{cc} {\bf Y}_1 & {\bf X} \\ -{\bf X}^{\dagger } & {\bf Y}_2
\end{array}
\right] ,
\label{XY-matrixA}
\end{equation}
with the definitions
$$
{\bf X} = \left[
\begin{array}{cc} 0 & v_+\\ -v_- & 0
\end{array}
\right]
\quad
 {\bf Y}_1 = \left[
\begin{array}{cc}
u_+ & 0\\ 0 & u_-
\end{array}
\right]
\quad
{\bf Y}_2 = \left[
\begin{array}{cc}
u_- & 0\\ 0 & u_+
\end{array}
\right],
$$
where the requirement of unitarity implies that
\begin{equation}
u_\pm^2+v_\pm^2=1.
\nonumber
\end{equation}
The $u$'s and $v$'s are variational parameters in the matrix
representation and are related to the parameters $\eta_{00}$ and
$\eta_{10}$ in Eq.\ (\ref{eq10b}).

The Bogoliubov-type transformation (\ref{XY-matrixA}) applied to
the $D$--$\pi$ pair space may be viewed as a quasiparticle
transformation that is further constrained to preserve the SU(4)
symmetry. The physical vacuum state $|0^*\rangle$ is transformed
to a quasiparticle vacuum state $|\psi\rangle$ and the basic
fermion operators,
$$
\{c^\dagger_{{\bf r}\uparrow},c^\dagger_{{\bf r}\downarrow},
c_{\bar{\bf r}\uparrow},c_{\bar{\bf r}\downarrow} \} ,
$$
are converted to quasifermion operators,
$$
\{a^\dagger_{{\bf r}\uparrow}, a^\dagger_{{\bf r}\downarrow},
a_{\bar{{\bf r}}\uparrow}, a_{\bar{{\bf r}}\downarrow} \}
\quad\mbox{with}\quad
a_{{\bf r}i}|\psi\rangle=0 ,
$$
through the transformation
\begin{equation} {\cal T}
\left (
\begin{array}{c} c_{{{\bf r}}\uparrow}
\\ c_{{{\bf r}}\downarrow}
\\ c^\dagger_{\bar{\bf r}\uparrow}
\\ c^\dagger_{\bar{\bf r}\downarrow}
\end{array}
\right) \left |0^*\right\rangle=
\left(
\begin{array}{c} a_{{{\bf r}}\uparrow}
\\ a_{{{\bf r}}\downarrow}
\\ a^\dagger_{\bar{\bf r}\uparrow}
\\ a^\dagger_{\bar{\bf r}\downarrow}
\end{array}
\right)\left |\psi\right\rangle.
\label{transfA}
\end{equation}

Using the transformation (\ref{transfA}) and the mapping
(\ref{mapping}), one can express any one-body operator in the
quasiparticle space as
\begin{eqnarray}
{\cal T}\hat O{\cal T}^{-1} &=&\left[
\begin{array}{cc} {\cal
O}^{(11)} & {\cal O}^{(12)} \\   {\cal
O}^{(21)} & {\cal O}^{(22)}
\end{array}\right]
\nonumber\\
\Rightarrow \hat{\cal O}&=& \sum'_{{\bf r},i} {\cal
O}^{(22)}_{ii}\nonumber +\sum'_{{\bf r},i,j} \left\{ {\cal
O}^{(11)}_{ij}a^\dagger_{{\bf r} i} a_{{\bf r} j} - {\cal
O}^{(22)}_{ji} a^\dagger_{\bar{{\bf r}} i} a_{\bar{{\bf r}}
j}\right.\nonumber\\ &+& \left.{\cal O}^{(12)}_{i,j}
a^\dagger_{{\bf r} i} a^\dagger_{\bar{{\bf r}} j} + {\cal
O}^{(21)}_{i,j} a_{\bar{{\bf r}} i} a_{{\bf r} j} \right\},
\label{transf2}
\end{eqnarray}
where we put a prime on the summation symbols to indicate that the
summation runs only over ${\bf r}\in$ even lattice sites. The
${\cal O}^{(\mu\nu)}_{ij}$ are fixed by the transformation
properties of the operator $\hat O$:
\begin{equation} {\cal O}^{(\mu\nu)}_{ij} =\sum_{m,n} [\ {\cal T}^{(\mu
m)}O^{(mn)} ({\cal T}^{-1})^{(n \nu)} ]_{ij},
\label{omn}
\end{equation}
and ${\cal T}^{(\mu m)}$ and $O^{(mn)}$ are two-dimensional
submatrices of ${\cal T}$ and $\hat O$, respectively.

Because the quasiparticle annihilation operator acting on the
quasiparticle vacuum $|\psi\rangle$ gives zero, the expectation
values for one-body operators $\hat{O}$ are given by
\begin{equation}
\langle\hat O\rangle = \langle\psi|\hat{\cal O}|\psi\rangle =
\sum'_{{\bf r},i} {\cal O}^{(22)}_{ii}=\sum'_{\bf r}{\rm Tr}
({\cal O}^{(22)}),
\label{onebody}
\end{equation}
and for two-body operators $\hat O_A \hat O_B$,
\begin{eqnarray}
\langle\hat O_A \hat O_B\rangle &=&
\langle\psi|\hat{\cal O_A}\hat{\cal O_B}|\psi\rangle \nonumber\\ &=&
\sum'_{\bf r}{\rm Tr} ({\cal O}^{(22)}_A) \sum'_{{\bf r}'}{\rm Tr} ({\cal
O}^{(22)}_B)
\label{twobody}
\\ &+&\sum'_{\bf r} {\rm Tr} ({\cal O}^{(21)}_A {\cal
O}^{(12)}_B).
\nonumber
\end{eqnarray}
Utilizing Eqs.\ (\ref{XY-matrixA}) and
(\ref{omn})--(\ref{twobody}), and noting that the summation
$\scriptstyle \sum'_{\bf r}$ provides a factor of $\Omega/2$
because the matrix elements of Eq.\ (\ref{omn}) do not depend on
${\bf r}$, one obtains the expectation values for the SU(4)
generators and their scalar products in the coherent state
representation.  For one-body terms,
\begin{eqnarray}
\langle D^{\dagger}\rangle &=&\langle D\rangle=-\tfrac \Omega2
(u_+v_+ +u_-v_-)
\nonumber
\\\langle \pi^{\dagger}_z\rangle &=&\langle \pi_z\rangle=-\tfrac
\Omega2 (u_+v_+ -u_-v_-)
\nonumber
\\  \langle {\cal Q}_z\rangle &=&
\tfrac \Omega2 (v_+^2 - v_-^2)
\label{eqoneA}
\\
\langle\hat n\rangle &=&\Omega (v_+^2 +
v_-^2)
\nonumber
\\
\langle \pi_x\rangle &=& \langle \pi_y\rangle=\langle
\vec{S}\rangle = \langle {\cal Q}_x\rangle=\langle {\cal
Q}_y\rangle=0 ,
\nonumber
\end{eqnarray}
and for two-body terms
\begin{eqnarray}
\langle D^{\dagger }D\rangle &=& \langle D\rangle^2 = \tfrac14
\Omega^2 (u_+v_+ + u_-v_-)^2
\nonumber
\\
\langle \vec{\pi}^{\dagger }\cdot\vec{\pi}\rangle &=& \langle
\pi_z\rangle^2 = \tfrac14 \Omega^2 (u_+v_+ - u_-v_-)^2
\nonumber
\\
\langle \vec{\cal Q} \cdot \vec{\cal Q}\rangle &=& \langle {\cal
Q}_z\rangle^2 = \tfrac14 \Omega^2 (v_+^2 - v_-^2)^2
\label{eqtwoA}
\\
\langle \vec{S} \cdot \vec{S}\rangle &=& 0,
\nonumber
\end{eqnarray}
where we have applied a large-$\Omega$ approximation, ignoring
terms that are of order $1/\Omega$ smaller than the leading terms.
Under this approximation, the expectation value for a two-body
operator is simply a product of the expectation values of two
one-body operators (that is, the last term on the right side of
Eq.\ (\ref{twobody}) is negligible).

\section{Derivation of the Temperature-Dependent Gap Equations}

The formulas in Appendix A are valid only at zero temperature.  In
this Appendix we extend the derivation to the finite temperature
case. Let us begin with a more general Hamiltonian, assuming that
the single-particle energies are degenerate only for adjacent
sites:
\begin{eqnarray}
H &=& \sum_{r={\rm even}}\varepsilon_{r}\hat{n}_{r} - \left( G_0
D^\dag D  +G_1\vec{\pi}^\dag\cdot\vec{\pi} \right.
\nonumber
\\
&& + \chi\vec{Q}\cdot\vec{Q}+\kappa\vec{S}\cdot\vec{S} ),
\label{hamb1}
\end{eqnarray}
where
\begin{equation}
\hat{n}_{r}\equiv c^\dagger_{{\bf r}\uparrow}c_{{\bf r}\uparrow}+
c^\dagger_{\bar{\bf r}\downarrow}c_{\bar{\bf r}\downarrow} +
c^\dagger_{{\bf r}\downarrow}c_{{\bf r}\downarrow}+
c^\dagger_{\bar{\bf r}\uparrow}c_{\bar{\bf r}\uparrow}.
\nonumber
\end{equation}
This Hamiltonian is not an invariant with respect to the SU(4)
symmetry generated by the operators of Eq.\ (\ref{rspace2}).
However, with a slight revision, the following 16 operators form
an U(4) algebra denoted as U$_4$(r) (with an SU$_4$(r) subgroup),
\begin{eqnarray}
p_{12}^\dagger({\bf r})&=& \left( c_{{\bf r}\uparrow}^\dagger
c^\dagger_{\bar{\bf r}\downarrow} -c_{{\bf r}\downarrow}^\dagger
c^\dagger_{\bar{\bf r}\uparrow}\right)
\nonumber
\\
q_{ij}^\dagger({\bf r})&=& \left( c_{{\bf r},i}^\dagger
c^\dagger_{\bar{\bf r},j}+c_{{\bf r},j}^\dagger
c^\dagger_{\bar{\bf r},i}\right)
\nonumber
\\
S_{ij}({\bf r}) &=& \left( c_{{\bf r},i}^\dagger c_{{\bf r},j}-
c_{\bar{\bf r},j} c_{\bar{\bf r},i}^\dagger\right)
\nonumber
\\
\tilde{Q}_{ij}({\bf r})&=& \left( c_{{\bf r},i}^\dagger
c_{{\bf r},j}+ c_{\bar{\bf r},j}c_{\bar{\bf r},i}^\dagger\right)
\nonumber
\\
p_{12}({\bf r}) &=& \left(p_{12}^\dagger({\bf r})\right)^\dagger
\qquad q_{ij}({\bf r}) = \left(q_{ij}^\dagger({\bf
r})\right)^\dagger
\nonumber
\end{eqnarray}
with
\begin{eqnarray}
p_{12}^\dagger&=& \sum_{r={\rm even}} p_{12}^\dagger({\bf r})
\qquad p_{12}=\sum_{r={\rm even}}   p_{12}({\bf r})\
\nonumber \\
q_{ij}^\dagger&=&\sum_{r={\rm even}} q_{ij}^\dagger({\bf r})
\qquad q_{ij}=\sum_{r={\rm even}} q_{ij}({\bf r})
\nonumber \\
\tilde{Q}_{ij}&=& \sum_{r={\rm even}}\tilde{Q}_{ij}({\bf r})
\qquad S_{ij} = \sum_{r={\rm even}}S_{ij}({\bf r}) .
\nonumber\
\end{eqnarray}
As for Eq.\ (\ref{operatorset}), one can also define more
physical operators $D({\bf r})$, $\vec{\pi}({\bf r})$,
$\vec{Q}({\bf r})$, $\vec{S}({\bf r})$, $M({\bf r})$, and
$\hat{n}_{r}=2M({\bf r})+2$, and express them as
\begin{eqnarray}
D &=& \sum_{r={\rm even}}D({\bf r}) \qquad \vec{\pi} =
\sum_{r={\rm even}}\vec{\pi}({\bf r})
\nonumber
\\
\vec{Q} &=& \sum_{r={\rm even}}\vec{Q}({\bf r}) \qquad \vec{S} =
\displaystyle \sum_{r={\rm even}}\vec{S}({\bf r})
\nonumber
\\
M &=& \sum_{r={\rm even}}M({\bf r}).
\nonumber
\end{eqnarray}
The Hamiltonian (\ref{hamb1}) has the symmetry of the direct
product of SU$_4$(r). The coherent state now becomes
\begin{eqnarray}
\mid \psi\rangle &=& \prod_{r={\rm even}} e^ { (\eta_{00}(r)
D^{\dagger }({\bf r})+\eta_{10}(r) {\pi}^{\dagger }_z({\bf
r})-{\rm h.\ c.}) } \mid 0^*\rangle
\nonumber
\\
&\equiv& {\cal T}\mid 0^*\rangle,
\nonumber
\end{eqnarray}
where ${\cal T}$ is a direct product of the ${\cal T}(r)$,
\begin{equation}
 {\cal T} = \prod_{r={\rm even}}{\cal T}({\bf r}) = \prod_{r={\rm even}}
\left[
\begin{array}{cc} {\bf Y}_1({\bf r}) & {\bf X}({\bf r}) \\
-{\bf X}^{\dagger}({\bf r}) & {\bf Y}_2({\bf r})
\end{array}
\right] ,
\nonumber
\end{equation}
with
\begin{eqnarray}
{\bf X}({\bf r}) &\equiv& \left[
\begin{array}{cc} 0 & v_{r+} \\ - v_{r-} & 0
\end{array}
\right]
\nonumber
\\
{\bf Y}_1({\bf r}) &\equiv& \left[
\begin{array}{cc} u_{r+} & 0\\ 0 & u_{r-}
\end{array}
\right]
\nonumber
\\
{\bf Y}_2({\bf r}) &\equiv& \left[
\begin{array}{cc} u_{r-} & 0\\ 0 & u_{r+}
\end{array}
\right] .
\nonumber
\end{eqnarray}
The variational Hamiltonian is
\begin{eqnarray}
  H' &=& H-\lambda\hat{n}
\nonumber\\
   &=& \sum_{r={\rm even}} (\varepsilon_{r}-\lambda)\hat{n}_r
\nonumber
\\
&-& \left( G_0 D^\dag D + G_1\vec{\pi}^\dag\cdot\vec{\pi}
+\chi\vec{Q}\cdot\vec{Q}+\kappa\vec{S}\cdot\vec{S} \right).
\nonumber
\end{eqnarray}
Taking expectation values of operators as in Eqs.\ (\ref{eqoneA})
and (\ref{eqtwoA}) permits the expectation value of $H'$ to be
determined. However, to consider the temperature dependence, one
should replace the quasiparticle vacuum state $|\psi\rangle$ by
the state $|\psi(T)\rangle$ in which quasiparticles are thermally
excited. The expectation values of $a^\dagger_{ri}a_{rj}$ and
$a^\dagger_{\bar{r} i}a_{\bar{r} j}$ in (\ref{transf2}) are no
longer zero when $i=j$.  Therefore, instead of (\ref{onebody}),
the formula to evaluate expectation values for a one-body operator
becomes
\begin{eqnarray}
\langle\hat O\rangle &=& \langle\psi(T)|\hat{\cal
O}|\psi(T)\rangle
\nonumber
\\
&=& \sum_{r={\rm even}} \left\{ {\cal O}^{(22)}_{11}(r)+{\cal
O}^{(22)}_{22}(r) \right.
\nonumber
\\
 &+& [ {\cal O}^{(11)}_{11}(r)\tilde{n}_{r\uparrow}(T)
-{\cal O}^{(22)}_{22}(r) \tilde{n}_{\bar{r}\downarrow}(T) ]
\nonumber
\\
&+& \left. [{\cal O}^{(11)}_{22}(r)\tilde{n}_{{r}\downarrow}(T)
-{\cal O}^{(22)}_{11}(r) \tilde{n}_{\bar{r}\uparrow}(T) ] \right\}
.
\label{onebodyexp}
\end{eqnarray}
In Eq.\ (\ref{onebodyexp}), we have supposed that at temperature
$T$, $\tilde{n}_{r i}(T)$ and $\tilde{n}_{\bar{r} i}(T)$ are
respectively the numbers of quasiparticles at $r$ and $\bar{r}$
with spin $i$,
\begin{eqnarray}
\langle\psi(T)|a^\dagger_{r i}a_{r i}|\psi(T)\rangle &=&
\tilde{n}_{ri}(T) \nonumber
\\
\langle\psi(T)|a^\dagger_{\bar{r} i}a_{\bar{r} i}|\psi(T)\rangle
&=& \tilde{n}_{\bar{r} i}(T). \nonumber
\end{eqnarray}
It can be shown that for all the SU(4) generators except spin
$\vec{S}$,
$$
{\cal O}^{(11)}_{11}(r)=-{\cal O}^{(22)}_{22}(r) \qquad {\cal
O}^{(11)}_{22}(r)=-{\cal O}^{(22)}_{11}(r) ,
$$
and therefore
\begin{eqnarray}
\langle\hat O\rangle &=& \langle\psi|\hat{\cal O}|\psi\rangle
\nonumber
\\
&=& \sum_{r={\rm even}} \left\{ {\cal O}^{(22)}_{11}(r)\ [\
1-\tilde{n}_{r-}(T) ] \right. \nonumber
\\
&+& \left. {\cal O}^{(22)}_{22}(r)\ [\ 1- \tilde{n}_{r+}(T)\ ]\
\right\} ,
\label{onebody2}
\end{eqnarray}
where
\begin{equation}
\tilde{n}_{r+}\equiv a^\dagger_{{\bf r}\uparrow}a_{{\bf
r}\uparrow}+ a^\dagger_{\bar{\bf r}\downarrow}a_{\bar{\bf
r}\downarrow},
\quad
\tilde{n}_{r-}\equiv a^\dagger_{{\bf
r}\downarrow}a_{{\bf r}\downarrow}+ a^\dagger_{\bar{\bf
r}\uparrow}a_{\bar{\bf r}\uparrow}.
\label{eq122}
\end{equation}
As for Eq.\ (\ref{eqoneA}), one can use Eq.\ (\ref{onebody2})
to obtain the one-body terms
\begin{subequations}
\label{onebodyT:whole}
\begin{eqnarray} \langle D^{\dagger}(r)\rangle
&=&\langle D(r)\rangle \nonumber
\\
&=& - \left[ P_{r+}(T)u_{r+}v_{r+} \right. \nonumber
\\
&& \left. + P_{r-}(T)u_{r-}v_{r-} \right] \label{eq123}
\\
\langle \pi^{\dagger}_z(r)\rangle&=&\langle \pi_z(r)\rangle
\nonumber
\\
&=& - \left[  P_{r+}(T)u_{r+}v_{r+} \right. \nonumber
\\
&& \left. - P_{r-}(T)u_{r-}v_{r-}  \right] \label{eq124}
\\
\langle \vec{Q}(r)\rangle &=&\langle{Q}_z(r)\rangle
\nonumber
\\
&=& \tfrac12 \left[ P_{r+}(T)(2v^2_{r+}-1) \right. \nonumber
\\
 &&\left. - P_{r-}(T)(2v^2_{r-}-1) \right] \label{eq125}
\\
\left \langle M(r) \right \rangle\, &=&\tfrac12 n_r-1
\nonumber
\\
&=& \tfrac12 \left[
             P_{r+}(T)(2v^2_{r+}-1) \right.
\nonumber
\\
&& \left. + P_{r-}(T)(2v^2_{r-}-1) \right] , \label{eq126}
\end{eqnarray}
\end{subequations}
where
\[ P_{r\pm}(T)= 1-\tilde{n}_{r\pm}(T).\]
As in Eqs.\ (\ref{eqoneA}), $\langle\pi_x\rangle$,
$\langle\pi_y\rangle$, $\langle \vec{Q}_x\rangle$, $\langle
\vec{Q}_y\rangle$, and $\langle\vec{S}\rangle $ are all zero. Note
that for spin this is true only when the quasiparticles result
from thermal excitations. This is because from (\ref{onebodyexp})
one can show that
\begin{eqnarray}
\langle \vec{S}(r)\rangle &=& \langle S_z(r)\rangle
\nonumber
\\
&=& \frac12\left\{(\tilde{n}_{r\uparrow}
-\tilde{n}_{\bar{r}\downarrow})
-(\tilde{n}_{r\downarrow}-\tilde{n}_{\bar{r}\uparrow})\right\},
\nonumber
\end{eqnarray}
which is generally nonzero.  Only when the quasiparticles are due
to thermal excitations, as we will show later, we do have
$\tilde{n}_{r\uparrow} =\tilde{n}_{\bar{r}\downarrow}$ and
$\tilde{n}_{r\downarrow}=\tilde{n}_{\bar{r}\uparrow}$, which leads
to $\langle S_z(r)\rangle=0$.

Applying the large-$\Omega$ approximation (ignoring the second
term in Eq.\ (\ref{twobody}) when $\Omega\rightarrow\infty$), one
obtains the two-body terms
\begin{subequations}
\label{twobodyT:whole}
\begin{eqnarray}
        \left \langle D^\dag D\right \rangle
 &=& \langle D\rangle^2
\nonumber
\\
&=&  \left\{ \sum_{r={\rm even}} \left[ P_{r+}(T)u_{r+}v_{r+}
\right. \right. \nonumber
\\
&& \left. \left. + P_{r-}(T)u_{r-}v_{r-} \right]
\vphantom{\sum_{r={\rm even}}} \right\}^2
\label{eq128}
\\
 \left \langle  \vec{\pi}^\dag\cdot\vec{\pi} \right \rangle
&=& \langle \pi_z\rangle^2
\nonumber
\\
&=& \left\{ \sum_{r={\rm even}} \left[ P_{r+}(T)u_{r+}v_{r+}
\right. \right. \nonumber
\\
&& \left. \left. - P_{r-}(T)u_{r-}v_{r-} \right]
\vphantom{\sum_{r={\rm even}}} \right\}^2
\\
 \left \langle \vec{Q}\cdot\vec{Q} \right \rangle&=&\ \langle
Q_z\rangle^2
\nonumber
\\
&=& \frac14
       \left\{ \sum_{r={\rm even}}  \left[ P_{r+}(T)
(v^2_{r+}-u^2_{r+}) \right. \right.
\nonumber
\\
&& \left. \left. -P_{r-}(T)(v^2_{r-}-u^2_{r-}) \right]
\vphantom{\sum_{r={\rm even}}} \right\}^2
\\
\left \langle\ M\ \right \rangle &=&-\frac 12x\Omega
\nonumber
\\
 &=&\frac12 \sum_{r={\rm even}}
            \left[ P_{r+}(T)(v^2_{r+}-u^2_{r+}) \right.
\nonumber
\\
&& \left. + P_{r-}(T)(v^2_{r-}-u^2_{r-})\right].
\label{eq131}
\end{eqnarray}
\end{subequations}
The next operations are similar to those of section III.C. By
introducing the energy gaps
$$
\Deltad=|G_0\langle D\rangle| \qquad \Deltapi=|G_0\langle
\pi_z\rangle| \qquad \Deltaq=| \chi\langle \vec{Q}_z\rangle|,
$$
defining $\varepsilon_{r\pm}$ and $\Deltapm $, and utilizing
(\ref{onebodyT:whole}) to construct $\langle H' \rangle$ and
perform the variation calculation $\delta\langle H' \rangle=0$,
one obtains
\begin{equation}
2u_{r\pm}v_{r\pm} (\varepsilon_{r
\pm}-\lambda)-\Deltapm(u^2_{r\pm} -v^2_{r\pm})=0.
\label{C22}
\end{equation}
Solving this equation gives
\begin{subequations}
\label{uvT:whole}
\begin{eqnarray}
 u^2_{r\pm} &=&\frac{1}{2}\left
[1+\frac{\varepsilon_{r\pm}-\lambda}{e_{r\pm}}\right]
\label{uequation}
\\
v^2_{r\pm} &=& \frac{1}{2}\left
[1-\frac{\varepsilon_{r\pm}-\lambda}{e_{r\pm}}   \right]
\label{vequation}
\end{eqnarray}
\end{subequations}
with quasiparticle energies defined as
\begin{equation}
 e_{r\pm}=\sqrt{(\varepsilon_{r\pm}-\lambda)^2+{\Deltapm}^2} .
\label{quasieT}
\end{equation}
Inserting Eqs.\ (\ref{uvT:whole}) into Eqs.\
(\ref{onebodyT:whole}) and (\ref{twobodyT:whole}), one finds the
gap equations:
\begin{subequations}
\label{gapsT:whole}
\begin{eqnarray}
        \Deltad &=& \frac{G_0}{2}\sum_{r={\rm even}}
\left[ \frac{P_{r+}(T)}{e_{r+}}
 \DeltaPlus \right.
\label{eq139a}
\nonumber
\\
&& \left. + \frac{P_{r-}(T)}{e_{r-}}\ \DeltaMinus \right]
\\
 \Deltapi &=& \frac{G_1}{2}\sum_{r={\rm even}}
\left[ \frac{P_{r+}(T)}{e_{r+}}
 \DeltaPlus \right.
\nonumber
\\
&& \left. - \frac{P_{r-}(T)}{e_{r-}}\
\DeltaMinus \right] \\
\Deltaq &=& \frac{\chi}{2}\sum_{r={\rm even}} \left[
\frac{P_{r+}(T)}{e_{r+}}
 (\lambda-\varepsilon_{r+})  \right.
\nonumber
\\
&& \left. - \frac{P_{r-}(T)}{e_{r-}}\
(\lambda-\varepsilon_{r-}) \right] \\
-2x &=& \frac{2}{\Omega}\,\sum_{r={\rm even}} \left[
\frac{P_{r+}(T)}{e_{r+}}
 (\lambda-\varepsilon_{r+}) \right.
\nonumber
\\
&& \left. + \frac{P_{r-}(T)}{e_{r-}}\ (\lambda-\varepsilon_{r-})
\right],
\label{eq139}
\end{eqnarray}
\end{subequations}
The energy of the system is then obtained as
\begin{eqnarray}
E(T)&=&\langle  H' \rangle +n\lambda\nonumber\\
&=&  \sum_r  n_r \varepsilon_{r} -\left [\ \frac{{\Deltad}^2}{
G_0} +\frac{{\Deltapi}^2}{ G_1}+\frac{{\Deltaq}^2}{ \chi}\
\right].
\nonumber
\end{eqnarray}

Utilizing Eqs.\ (\ref{uvT:whole})--(\ref{gapsT:whole}), it can be
shown that
\begin{eqnarray*}
&-& 2\left [\frac{{\Deltad}^2}{G_0} +\frac{{\Deltapi}^2}{G_1}
+\frac{{\Deltaq}^2}{\chi} \right]_{T=0} +\sum_{r={\rm
even}}(\varepsilon_r-\lambda) n_r
\nonumber \\
&=& \sum_{r={\rm even}}\left[ 2(\varepsilon_r-\lambda) -
P_{r+}(T)e_{r+}- P_{r+}(T)e_{r-}
 \right].
\end{eqnarray*}
This relation can be used to rewrite the energy as a sum of the
quasiparticle vacuum energy and the quasiparticle excitation
energy
\begin{equation}
E(T)=E_0(T)+\sum_{r={\rm even}} \left\{\ \tilde{n}_{r+}(T)\ e_{r+}
+ \tilde{n}_{r-}(T)\ e_{r-}\ \right\} .
\label{C32}
\end{equation}
In the above equation, the quasiparticle vacuum energy is
\begin{eqnarray}
E_0(T) &=& \sum_{r={\rm even}}
(2\varepsilon_{r}-e_{r+}- e_{r-})-(\Omega-n)\lambda
\nonumber
\\
 &+& \left [\ \frac{{\Deltad}^2}{G_0}
+\frac{{\Deltapi}^2}{G_1}
+\frac{{\Deltaq}^2}{\chi} \right]_{T=0} ,
\label{C33}
\end{eqnarray}
which is determined by solving the gap equations
(\ref{gapsT:whole}) at $T=0$, while the quasiparticle number
operators are assumed to be
\begin{equation}
\tilde{n}_{r\pm}(T)=\frac{2}{1+\exp \left(\dfrac{e_{r\pm}}{k\tsub
B T}\right)} ,
\nonumber
\end{equation}
since they are associated with thermal fluctuations, and thus
\begin{equation}
P_{r\pm}(T)= 1-\tilde{n}_{r\pm}(T) = \tanh
\left(\frac{e_{r\pm}}{2k\tsub B T} \right) .
\nonumber
\end{equation}
Note that, according to the definition (\ref{eq122}),
\begin{eqnarray*}
\tilde{n}_{r+}(T) &=&
\tilde{n}_{r\uparrow}(T)+\tilde{n}_{\bar{r}\downarrow}(T)
\\
\tilde{n}_{r-}(T) &=&
\tilde{n}_{r\downarrow}(T)+\tilde{n}_{\bar{r}\uparrow}(T).
\end{eqnarray*}
Therefore, Eq.\ (\ref{C32}) implies that the quasiparticle states
$|{r\uparrow}\rangle$ and $|{\bar{r}\downarrow}\rangle$ are
degenerate with excitation energy $e_{r+}$, while the states
$|{r\downarrow}\rangle$ and $|{\bar{r}\uparrow}\rangle$ are
degenerate with excitation energy $e_{r-}$. Thus, as we have
mentioned before,
\begin{eqnarray}
\tilde{n}_{r\uparrow}(T)&=&
\tilde{n}_{\bar{r}\downarrow}(T)=\frac{1}{1+
\exp\left(\dfrac{e_{r+}}{k\tsub B T}\right)}
\nonumber\\
\tilde{n}_{r\downarrow}(T)&=&\tilde{n}_{\bar{r}\uparrow}(T)=\frac{1}{1+
\exp \left(\dfrac{e_{r-}}{k\tsub B T}\right)} .
\nonumber
\end{eqnarray}
The physical meaning of $\Deltaq$ can now be better understood.
From the expression for the quasiparticle energies $e_{r\pm}$, one
sees that the original four-fold degeneracy of the single-particle
energy level $\varepsilon_{r}$ has been split into a two-fold
degeneracy because of the $\vec{Q}\cdot\vec{Q}$ interaction:
\begin{eqnarray*}
\varepsilon_{r+} &=& \varepsilon_{r}-\Deltaq
\qquad
(|{r\uparrow}\rangle {\rm\ and\ } |{\bar{r}\downarrow}\rangle
\\
\varepsilon_{r-} &=& \varepsilon_{r}+\Deltaq
\qquad
(|{r\downarrow}\rangle {\rm\ and\ }
|{\bar{r}\uparrow}\rangle.
\end{eqnarray*}
The energy difference $2\Deltaq$ is just the energy required to
flip the spin of an electron from $|{r\uparrow}\rangle$ to
$|{r\downarrow}\rangle$ or from $|{\bar{r}\uparrow}\rangle$ to
$|{\bar{r}\downarrow}\rangle$, and {\em vice versa}.

The above results can also be obtained through a mean field
approximation. Note that any two-body operator of the form
$\hat{O}_A\hat{O}_B$ can always be written as
\begin{eqnarray*}
\hat{O}_A\hat{O}_B&=&
[\langle\hat{O}_A\rangle +(\hat{O}_A-\langle\hat{O}_A\rangle)]
\nonumber
\\
&& \times ~ [\langle\hat{O}_B\rangle
+(\hat{O}_B-\langle\hat{O}_B\rangle)]
\nonumber \\
 &=&
\langle\hat{O}_A\rangle\langle\hat{O}_B\rangle
+\hat{O}_A(\hat{O}_B-\langle\hat{O}_B\rangle)
\nonumber
\\
&& + ~ (\hat{O}_A-\langle\hat{O}_A\rangle)\hat{O}_B\\
&& + ~
(\hat{O}_A-\langle\hat{O}_A\rangle)(\hat{O}_B-\langle\hat{O}_B\rangle).
\nonumber
\end{eqnarray*}
Suppose that the last term of the preceding equation can be
ignored, implying that the physical effects of
$(\hat{O}_A-\langle\hat{O}_A\rangle)$ and
$(\hat{O}_B-\langle\hat{O}_B\rangle)$ are small. Then the two-body
operator reduces to an one-body operator:
\begin{equation}
\hat{O}_A\hat{O}_B= -\langle\hat{O}_A\rangle\langle\hat{O}_B\rangle
+\hat{O}_A\langle\hat{O}_B\rangle
+\langle\hat{O}_A\rangle\hat{O}_B .
\nonumber
\end{equation}
Applying this equation to the Hamiltonian (\ref{hamb1}),
\begin{equation}
H = H_0+\sum_{r={\rm even}}h(r) ,
\nonumber
\end{equation}
where $H_0$ is a c-number
\begin{eqnarray}
H_0&=&\sum_{r= {\rm even}} 2\varepsilon_{r}-(\Omega-n)\lambda
\nonumber
\\
&&  + \left [\frac{{\Deltad}^2}{G_0}
+\frac{{\Deltapi}^2}{G_1}
+\frac{{\Deltaq}^2}{\chi} \right]_{0},
\label{C40}
\end{eqnarray}
the subscript $0$ denotes a ground state expectation value, we
have made the replacements
$$
-G_0\langle D\rangle \rightarrow \Deltad,
\quad
-G_1\langle \pi_z\rangle \rightarrow \Deltapi,
\quad
\chi\langle Q_z\rangle \rightarrow \Deltaq,
$$
and the second term in $H$ defines a mean field
\begin{eqnarray}
h(r)&=& 2(\varepsilon_r-\lambda)M(r)+\left\{
\Deltad [D^\dagger(r)+D(r)] \right. \nonumber
\\
&& \left. +\Deltapi [\pi_z^\dagger(r)+\pi_z(r)]
-2\Deltaq Q_z \right\} .
\label{C41}
\end{eqnarray}
Now we implement a quasiparticle transformation on $H$,
\begin{equation}
{\cal H}\equiv {\cal T} H {\cal T}^{-1}=H_0+\sum_{r={\rm even}}
\tilde{h}(r),
\nonumber
\end{equation}
where
\begin{equation}
\tilde{h}(r)\equiv {\cal T} h(r) {\cal T}^{-1}=\left(
\begin{array}{cc}
\tilde{h}^{(11)}(r)& \tilde{h}^{(12)}(r)\\
\tilde{h}^{(21)}(r)& \tilde{h}^{(22)}(r)
\end{array}\right)
\nonumber
\end{equation}
and
\begin{eqnarray}
\tilde{h}^{(11)}(r) &=&\left(
\begin{array}{cc}
e_{r+}& 0\\ 0 & e_{r-}
\end{array}\right)
\nonumber
\\
\tilde{h}^{(22)}(r) &=& \left(
\begin{array}{cc}
-e_{r-}& 0\\ 0 & -e_{r+}
\end{array}\right)
\nonumber
\\
\tilde{h}^{(12)}(r) &=& \tilde{h}^{(21)}(r)^\dagger=\left(
\begin{array}{cc}
0 & {\cal O}_+\\
-{\cal O}_-& 0
\end{array}\right),
\nonumber
\end{eqnarray}
with
\begin{eqnarray}
{\cal O}_\pm&=&\Deltapm (u_{r\pm}^2-v_{r\pm}^2)
\nonumber\\
&&- ~ (\varepsilon_{r\pm}-\lambda)\ 2u_{r\pm}v_{r\pm}, \label{C46}
\\
e_{r\pm}&=&(\varepsilon_{r\pm}-\lambda) (u_{r\pm}^2-v_{r\pm}^2)
\nonumber\\
&&+ ~ \Deltapm 2u_{r\pm}v_{r\pm} . \label{C47}
\end{eqnarray}
The condition for the Hamiltonian ${\cal H}$ to be diagonal is
that Eq.\ (\ref{C46}) be zero, which is equivalent to Eq.\
(\ref{C22}). The solutions of $u_{r\pm}$ and $v_{r\pm}$ are the
same as those in Eqs.\ (\ref{uvT:whole}). Inserting these $u$'s
and $v$'s into Eq.\ (\ref{C47}), one can check that the $e_{r\pm}$
appearing in (\ref{gapsT:whole}) are indeed the quasiparticle
energies defined in (\ref{quasieT}).

In the expression for an operator [see Eq.\ (\ref{mapping})],
$\tilde{h}(r)$ can be written as
\begin{eqnarray}
\tilde{h}(r)&=&e_{r+}a^\dagger_{r\uparrow}a_{r\uparrow}
+e_{r-}a^\dagger_{r\downarrow}a_{r\downarrow}
\nonumber
\\
&& -e_{r-}a_{\bar{r}\uparrow}a^\dagger_{\bar{r}\uparrow}
-e_{r+}a^\dagger_{\bar{r}\downarrow}a_{\bar{r}\downarrow}
\nonumber
\\
&=&-(e_{r+}+e_{r-})+(\tilde{n}_{r+}e_{r+} + \tilde{n}_{r-}e_{r-})
.\nonumber
\end{eqnarray}
Since
$$
H_0-\sum_r(e_{r+}+e_{r-})=E_0
$$
[see (\ref{C33}) and (\ref{C40})], the Hamiltonian may be written
as
\begin{equation}
{\cal H}=E_0+\sum_{r={\rm even}}[\tilde{n}_{r+}e_{r+} +
\tilde{n}_{r-}e_{r-}],
\end{equation}
which is Eq.\ (\ref{C32}). If we ignore single-particle energy
splitting, $u$ and $v$ have no $r$ dependence, meaning that
$\sum_r=\Omega/2$, and in this case all results reduce to those
discussed in Section III.

From the preceding discussion it becomes clear that the SU(4)
coherent state is in fact a mean field solution of the SU(4)
Hamiltonian. The SU(4) coherent state may thus be viewed as a
generalization of the BCS theory, and the process to derive it can
be characterized as a Hartree--Fock--Bogoliubov (HFB)
transformation that implements a (non-abelian) symmetry constraint
on the variational wavefunction.  It is a generalized BCS theory
because the system contains two kinds of quasiparticles instead of
one. In addition, the $\vec{Q}\cdot\vec{Q}$ interaction splits the
single-particle energy levels into two sets,
$$
\varepsilon_{r\pm}= \varepsilon_{r}\mp \Deltaq.
$$
These new features are primarily responsible for the differences
between high-$T\tsub c$ superconductivity and conventional
superconductivity described by the ordinary BCS theory.

\section{General Solutions of the Gap Equations}

The temperature-dependent gap equations are coupled algebraic
equations. In this section we present one way to solve these
equations. First, we rewrite Eqs.\ (\ref{subgapeq:d}) and
(\ref{subgapeq:pi}) in the form
\begin{eqnarray}
&& \left (\omega_0-w_+ \right )\Deltad +
     \left (\omega_1-w_+\right )
      \Deltapi=0 \nonumber
\label{A2.1}\\
&& \left (\omega_0-w_- \right )\Deltad -
     \left (\omega_1-w_-\right )
      \Deltapi=0 \nonumber
\label{A2.2}
\end{eqnarray}
with
\begin{equation}
\omega_0=\frac{2}{G_0\Omega} \qquad \omega_1=\frac{2}{G_1\Omega} .
\nonumber \label{A2.3}
\end{equation}
The condition for these coupled equations to have solutions is
\begin{equation}
\left   (\omega_0-w_+\right ) \left
(\omega_1-w_-\right) +\left (\omega_1-w_+\right )
\left   (\omega_0-w_-\right) =0. \nonumber
\label{A2.4}
\end{equation}
One then obtains
\begin{equation}
w_\pm=\bar{\omega}\,
\frac{w_\mp-\tilde{\omega}}{w_\mp-\bar{\omega}} \qquad
\bar{\omega}=\frac{\omega_0+\omega_1}{2} \qquad
\tilde{\omega}=\frac{\omega_0\omega_1}{\bar{\omega}}
\label{A2.5}
\end{equation}
and
\begin{equation}
\begin{array}{c}
\displaystyle \Deltapi =\left (\frac{w_-
-\omega_0}{w_--\omega_1}\right )\Deltad
\\[1.2em]
\displaystyle \Deltapi =-\left (\frac{w_+
-\omega_0}{w_+-\omega_1}\right )\Deltad .
\label{A2.6}
\end{array}
\end{equation}
By using Eqs.\ (\ref{gap:whole}) and the new equations
(\ref{A2.5}) and (\ref{A2.6}), a formal solution in terms of
$w_\pm$ can be expressed as
\begin{subequations}
\label{gapsTC:whole}
\begin{eqnarray}
\Deltad
&=&G_0\Omega\sqrt{\frac{(I-x^2)/4-(\Deltaq/\chi\Omega)^2}{1
+\left(\dfrac{1-w^{}_+/\omega_0} {1-w_+/\omega_1}\right )^2}}
\label{A2.9}
\\
\displaystyle \Deltapi &=& \displaystyle
G_1\Omega\sqrt{\frac{(I-x^2)/4-(\Deltaq/\chi\Omega)^2}{1
+\left(\dfrac{1-w_+/\omega_1} {1-w_+/\omega_0}\right )^2}}
\label{A2.10}
\\
\Deltaq &=&\frac {\chi}{2}\Omega\left | \frac{x(w_+-w_-)}
{\chi\Omega w_+w_- -(w_++w_-)} \right |
\label{A2.7}
\\
\lambdaPrime  &=& -\frac {x}{2}\left |
\frac{\chi\Omega(w_++w_-)-4} {\chi\Omega w_+w_- -(w_++w_-)} \right
| ,
\label{A2.8}
\end{eqnarray}
\end{subequations}
where we have made use of the SU(4) invariant [see
Eq.\ (\ref{su4const})]
\begin{equation}
\langle{\cal E}\tsub{SU4}\rangle= \frac{\Omega^2}4 (I-x^2) \qquad
I=(1-u)^2, \label{eqEsu4_u}
\end{equation}
and $u$ is the unpaired number density.

Eqs.\ (\ref{gapsTC:whole}) represent a formal solution only since
$w_\pm$ must be determined from Eq.\ (\ref{wpm}) in a
self-consistent manner. This can be done by by combining Eqs.\
(\ref{wpm}), (\ref{A2.5}), and (\ref{gapsTC:whole}). It turns out
that $w_+$ and $w_-$ satisfy the same equation. Adopting the
notation $w_\pm \equiv w$, we have
\begin{equation}
y^2=
\frac{\omega_0\omega_1[w^2(\omega_0+\omega_1-2\omega_q)
-2w\omega_0\omega_1 +2\omega_0\omega_1\omega_q]^2}
{w^2[w(\omega_0+\omega_1)-2\omega_0\omega_1]^2
(\omega_0-2\omega_q)(\omega_1-2\omega_q)}
\label{A2.11}
\end{equation}
with
$$
\omega_q = \frac{1}{\chi\Omega}.
$$
and for the left side of Eq.\ (\ref{A2.11})
\begin{eqnarray}
y&\equiv& y_\pm=\frac {x}{\sqrt{I[1 \pm \delta_{\pm} \Gamma(y)]}}\
, \label{eqy}
\end{eqnarray}
with
\begin{eqnarray}
\delta_{\pm} &=&1-\frac{P_\pm(T)^2}I
\nonumber
\\
&=& 1-
\frac{\tanh^2(
e_\pm / 2k\tsub BT)
}I
\label{eqdelta}\\
\Gamma(y)&=& \left|1+\frac{w^2(\omega_0-\omega_1)^2}
{2\omega_0\omega_1(\omega-\omega_0)(\omega-\omega_1)}\right | .
\label{eqgamma}
\end{eqnarray}
Eqs.\ (\ref{gapsTC:whole}) now become
\begin{subequations}
\label{gapsTC1:whole}
\begin{eqnarray}
\Deltad &=&\frac{|(\omega_1-w)|\sqrt{1-x^2-(2\omega_q\Deltaq)^2}}
{\sqrt{\omega_0^2(w-\omega_1)^2+\omega_1^2(w-\omega_0)^2}}
\label{A2.14}\\
\Deltapi &=&\frac{|(w-\omega_0)|\sqrt{1-x^2-(2\omega_q\Deltaq)^2}}
{\sqrt{\omega_0^2(w-\omega_1)^2+\omega_1^2(w-\omega_0)^2}}
\label{A2.15}\\
\Deltaq &=& \frac x{2}\left| \frac{(w-\omega_0)(w-\omega_1)}
{w^2(\bar{\omega}-\omega_q)-w\omega_0\omega+\omega_q\omega_0\omega_1}\right|
\label{A2.12}\\
\lambdaPrime  &=& -\frac {x}{2}\left | \frac{(w^2)-4\omega_q
w+(4\omega_q\bar{\omega}-\omega_0\omega_1)}
{w^2(\bar{\omega}-\omega_q)-w\omega_0\omega+\omega_q\omega_0\omega_1}\right|
\label{A2.13}
\end{eqnarray}
\end{subequations}

By introducing the variables
\begin{equation}
q_0=1-\frac{2\omega_q}{\omega_0} \qquad
q_1=1-\frac{2\omega_q}{\omega_1} \qquad
\bar{q}=1-\frac{\omega_q}{\bar{\omega}}
\nonumber\label{A2.16}
\end{equation}
and
\begin{equation}
x_q^2=\frac{\chi-G_0}{\chi-G_1}=\frac{q_0}{q_1},
\label{A2.17}
\end{equation}
Eq.\ (\ref{A2.11}) can be greatly simplified:
\begin{equation}
\frac y{x_q}=\pm\ \frac{\bar{q}w^2-\tilde{\omega}w+\tilde{\omega}}
{q_0 w(w-\tilde{\omega})}.
\label{A2.18}
\end{equation}
The signs ``$\pm$" are for hole doping ($x, y >0$) and electron
doping ($x, y <0$), respectively. Obviously, this means that the
solution has particle-hole symmetry: $w$ depends only on the
absolute value of $y$ (or $x$). For this reason, we shall regard
$y$ as the absolute value $|y|$ and ignore the $\pm$ sign in the
following discussions. Solving the quadratic equation
(\ref{A2.18}) gives
\begin{eqnarray}
{w}&=&\frac 2{\epsilon(y)}\ ,
\label{A2.19}\\
\epsilon(y)&=&\left(\chi-(\chi-G_0)\frac {y}{x_q}\right)\nonumber\\
&-&\eta\ (\chi-G_1)\sqrt{x_q(x_q-y)(1-x_qy)},
\label{A2.20}
\end{eqnarray}
where $\eta=\pm1$. Eqs.\ (\ref{A2.19}) and (\ref{A2.20}) determine
the value of $w$ and thus of $w_{\pm}$. Since $w_+ \ge{w}_-$, in
Eq.\ (\ref{A2.20}), $\eta=+1$ for $w_+$ and $\eta=-1$ for $w_-$.

Inserting Eq.\ (\ref{A2.20}) into Eqs.\ (\ref{eqgamma}) and
(\ref{gapsTC1:whole}) gives
\begin{equation}
\Gamma(y)= \left|1+\frac{(G_0-G_1)^2}{2[G_0-\epsilon(y)][G_1-\epsilon(y)]}\right|
\label{eqgamma2}
\end{equation}
\begin{equation}
\begin{array}{rclrcl}
\Deltaq &=& \displaystyle\frac{\chi}{2}\Omega q(y)
\qquad &
\Deltapi
&=&\displaystyle\frac{G_1}2\Omega T(y)
\\[0.8em]
\Deltad &=&\displaystyle\frac{G_0}2\Omega S(y)
\qquad &
\lambdaPrime
&=&-\displaystyle\frac{\gamma (y)}{2}\Omega x
\label{A2.22}
\end{array}
\end{equation}
with
\begin{subequations}
\label{gapsTC2:whole}
\begin{eqnarray}
&&\hspace{-28pt}q(y)=\frac{[\epsilon(y)-G_0][\epsilon(y)-G_1]\ x}
{\epsilon(y)^2-[2\epsilon(y)-G_0-G_1]\chi-G_0G_1}
\label{A2.23}\\
&&\hspace{-28pt}T(x)=[\epsilon(y)-G_1]\sqrt{\frac{I-x^2-q(y)^2}
{[G_0-\epsilon(y)]^2+[G_1-\epsilon(y)]^2}}
\label{A2.26}\\
&&\hspace{-28pt}S(y)= [G_0-\epsilon(y)]\sqrt{\frac{I-x^2-q(y)^2}
{[G_0-\epsilon(y)]^2+[G_1-\epsilon(y)]^2}}
\label{A2.25}\\
&&\hspace{-28pt}\gamma(x)=\frac{\epsilon(y)^2(G_0+G_1-\chi)
+G_0G_1[\chi-2\epsilon(y)]}
{\epsilon(y)^2-[2\epsilon(y)-G_0-G_1]\chi-G_0G_1} .
\label{A2.24}
\end{eqnarray}
\end{subequations}

Eqs.\ (\ref{gapsTC2:whole}) can be further simplified by using Eq.\
(\ref{A2.17}) and recognizing that
\begin{eqnarray}
\epsilon(y)-G_0&=&-(\chi-G_1)\ (\eta ab-a^2)
\nonumber\label{A2.27}\\
\epsilon(y)-G_1&=& (\chi-G_1)\ (b^2-\eta ab)
\nonumber\label{A2.28}\\
\epsilon(y)-\chi\ &=&-(\chi-G_1)\ (x_qy+\eta ab)
\nonumber\label{A2.29}
\end{eqnarray}
where
\begin{equation}
a\equiv\sqrt{x_q(x_q-y)} \qquad b\equiv\sqrt{1-x_qy} .
\label{A2.30}
\end{equation}
In terms of $a$ and $b$, Eqs.\ (\ref{eqgamma2}) and
(\ref{gapsTC2:whole}) can be rewritten as
\begin{eqnarray}
&&\Gamma(y)=\frac{a^2+b^2}{2ab} \nonumber\\
&&q(y)=\frac{abx}{x_q y} \nonumber\\
&&T(y)=a\ \sqrt{\frac{(I-x^2)\ x_q^2-a^2b^2}
{(a^2+b^2)\ x_q^2}} \nonumber\\
&&S(y)=b\ \sqrt{\frac{(I-x^2)\ x_q^2-a^2b^2} {(a^2+b^2)\ x_q^2} }
\nonumber \label{A2.33}\\
&&\gamma(y)=(\chi-G_1) \frac{a^2b^2-(x_qy)^2} {x_qy}+\chi .
\nonumber \label{A2.32}
\end{eqnarray}
Using Eq.\ (\ref{A2.30}) to convert $a$ and $b$ back to $y$
yields
\begin{eqnarray}
&&\Gamma(y)=\frac{(1-y^2)+(x_q-y)^2}{2\sqrt{x_q(x_q-y)(1-x_qy)}}
\nonumber\label{Gamma_y}\\
&&q(y)=\sqrt{(x_q-y)(x_q^{-1}-y)}\ \frac xy
\nonumber\label{q_y}\\
&&T(y)=\sqrt{x(x_q-y)}\ g(y)\
\nonumber\label{T_y}\\
&&S(y)=\sqrt{x(x_q^{-1}-y)}\ g(y)\
\nonumber\label{S_y}\\
&&\gamma(y)=(\chi-G_1)\frac{x_q(1-x_qy)}{y} +G_1
\nonumber\label{gamma_y}\\
&&g(y)=\sqrt{\frac xy+\frac{x_q(Iy^2-x^2)}{xy^2(1-2x_qy+x_q^2)}} .
\nonumber
\end{eqnarray}
Inserting the above results into Eqs.\ (\ref{A2.22}) and taking
into account the definition of $x_q$ in Eq.\ (\ref{A2.17}) gives
Eqs.\ (\ref{Tgap:whole}) in Sect.\ V.A.
\begin{eqnarray}
\Deltaq&=&\frac {\chi\Omega}{2}\sqrt{(x_q^{-1}-y)(x_q-y)}\ \frac
xy \nonumber\label{Dqfinal}
\\
\Deltad&=& \frac{G_0\Omega}{2} \sqrt {x(x_q^{-1}-y)}\, g(y)
\nonumber\label{Ddfinal}
\\
\Deltapi&=& \frac{G_1\Omega}{2} \sqrt {x(x_q-y)} \, g(y)
\nonumber\label{Dpifinal}
\\
\lambdaPrime\ &=&\frac{(G_1 - \chi)\Omega}{2} x_q \left( \frac xy
- x_qx \right) - \frac{G_1\Omega}2 x .
\nonumber\label{Lfinal}
\end{eqnarray}

For $T=0$, from Eq.\ (\ref{eqy}), one obtains immediately
$$
y=y_\pm=x
$$
because $\delta_\pm\rightarrow 0$. We thus have the results of
Eqs.\ (\ref{gapT0:whole}) discussed in Sect. IV.A. For $T>0$,
however, the self-consistency condition $y=y_+=y_-$ must be
satisfied, which requires that
$$
\sqrt{I[1+\delta_+ \Gamma(y)]}=\sqrt{I[1-\delta_-\Gamma(y)]}
$$
and thus that $ \delta_+  + \delta_- = 0 . $ According to Eq.\
(\ref{eqdelta}), this can be fulfilled only if
\begin{equation}
I=\frac {\tanh^2( e_+ / 2k\tsub BT )+\tanh^2( e_- / 2k\tsub BT
)}{2} , \nonumber
\end{equation}
which implies the existence of unpaired particles. The density of
unpaired particles can be deduced from Eq.\ (\ref{eqEsu4_u}):
$$
(1-u)=\sqrt{ \frac
{\tanh^2 \left(\dfrac{e_+}{2k\tsub BT}\right)
+\tanh^2\left(\dfrac{e_-}{2k\tsub BT}\right)}{2} }.
$$
This completes the derivation. Methods to obtain the actual gap
values for given $x$ and $T$ from the above equations have been
discussed extensively in Section V.

\bibliographystyle{unsrt}

\begin{thebibliography} {99}
\bibitem{particle-hole} The
particle--hole symmetry intrinsic to the SU(4) model does not mean
that hole-doped and electron-doped cuprates are expected to behave
in the same manner. Although the states of the model are
particle--hole symmetric, the interactions entering the effective
Hamiltonian would not be expected to be the same for hole-doped
and particle-doped compounds.  We shall show in future
publications that electron-doped cuprates are predicted to behave
very differently than hole-doped cuprates in the SU(4) model.  One
consequence is a prediction that it is difficult to obtain
superconductivity in electron-doped cuprates.
\bibitem{coherent-parameters} Generally, the term $\eta_{10}q_{12}^{\dagger}$
in Eq.\ (\ref{eq10b}) should be
$\vec{\eta}_1\cdot\vec{\pi}^{\dagger}$, and the parameters could
be complex as well. Thus the most general SU(4) coherent state
depends on eight real variables. The reduction of the coherent
state parameters to only two in Eq.\ (\ref{eq10b}) follows from
requiring time reversal symmetry and assuming conservation of spin
projection $S_z$  for the wavefunction.
\bibitem{SCAFNote}The different doping dependences of antiferromagnetic
and superconducting (pairing) order implied by the SU(4) symmetry
have been discussed in detail in Ref.\ \cite{Gu99}, Section VII.
There we show that the SU(4) symmetry requires that SC correlation
energy decrease slower than the AF correlation energy as hole
doping increases, implying that pairing inevitably wins with
sufficient hole doping. The gap equations introduced in this paper
represent a quantitative description of this competition.
\end{thebibliography}

\end{document}